\documentclass{aa}
\usepackage{graphicx,times}

\begin{document}

\title{X-ray spectroscopy of NGC~5548}

\author{ J.S. Kaastra \inst{1}
         \and
         K.C. Steenbrugge \inst{1}
         \and
         A.J.J. Raassen \inst{1},\inst{2}
         \and
         R.L.J. van der Meer \inst{1}
         \and
         A.C. Brinkman \inst{1}
         \and
         D.A. Liedahl \inst{3}
         \and
         E. Behar \inst{4}
         \and
         A. de Rosa \inst{5}
         }
  
\offprints{J.S. Kaastra}
\mail{J.Kaastra@sron.nl}

\institute{ SRON National Institute for Space Research
              Sorbonnelaan 2, 3584 CA Utrecht, The Nether\-lands 
              \and
              Astronomical Institute "Anton Pannekoek", Kruislaan 403,
              1098 SJ Amsterdam, The Netherlands
              \and
              Physics Department,
              Lawrence Livermore National Laboratory, 
              P.O. Box 808, L-11, Livermore, CA 94550, USA
              \and
              Department of Physics, 550 West 120th Street, New York, NY 10027, USA
              \and
              Istituto di Astrofisica Spaziale, CNR, 
              Via del Fosso del Cavaliere 100, Roma I-00133, Italy
              }

\date{Received  / Accepted  }

\abstract{
We analyze the high-resolution X-ray spectrum of the Seyfert 1 galaxy NGC~5548,
for the full $0.1$--$10$~keV band, using improved calibration results of the
Chandra-LETGS instrument.  The warm absorber consists of at least three
ionization components, namely one with a low, medium and high ionization
parameter.  The X-ray absorbing material, from an outflowing wind, covers the
full range of velocity components found from UV absorption lines.  The presence
of redshifted emission components for the strongest blue-shifted resonance
absorption lines indicate that the absorber is located at a distance larger than
the edge of the accretion disk.  We derive an upper limit to the edge of the
accretion disk of 1 light year.  Absorption lines from ions of at least ten
chemical elements have been detected, and in general for these elements there
are no strong deviations from solar abundances.  The narrow emission lines from
the \ion{O}{vii} and \ion{Ne}{ix} forbidden and intercombination lines probably
originate from much larger distances to the black hole.  We find evidence for
weak relativistically broadened oxygen and nitrogen emission lines from the
inner parts of the accretion disk, but at a much smaller flux level than those
observed in some other active galactic nuclei.  In addition, there is a broad,
non-relativistic \ion{C}{vi} Ly$\alpha$ emission line that is consistent with
emission lines from the inner part of the optical/UV broad line region.
\keywords{Galaxies: individual: NGC~5548 --
Galaxies: Seyfert -- quasars: absorption lines -- quasars: emission lines --
-- X-rays: galaxies }}

\maketitle

\section{Introduction}

Active Galactic Nuclei (AGN) show a rather violent environment.  Gas is being
swallowed by the black hole, which is fed by a continuous supply of fresh
material through the accretion disk.  This process becomes visible as intense
high-energy radiation from the inner parts of the disk and the immediate
surroundings of the black hole.  This radiation field may drive outflows from
the nucleus.  The detection of these outflows allows us to probe into the inner
nuclear regions.

At least 50~\% of Seyfert 1 galaxies exhibit signatures of photo-ionized gas in
their X-ray spectra (the so-called "warm absorber", e.g.  Reynolds
\cite{reynolds}; George et al.  \cite{george1998}).  In previous X-ray (e.g.,
ASCA) spectra the most prominent features were identified as O~VII and O~VIII
absorption edges.  However, these spectra did not have the energy resolution to
determine velocity shifts smaller than $\sim 3000$~km/s, or to perform any line
diagnostics.  Until recently, the only information regarding the dynamics of
circum-nuclear material along the line of sight has been from the detection of
UV absorption features.  These invariably indicate outflows with velocities of
hundreds of km/s (e.g.  Crenshaw et al.  \cite{crenshaw}).  Despite having a
high geometrical covering factor, the origin, location, velocity field, and
ionization structure of the X-ray absorber/emitter remained poorly understood
for nearly two decades.  In particular the relationship between the X-ray and UV
absorbers has presented considerable problems for formulating a consistent
picture of the inner nucleus in active galaxies.  Only recently sufficiently
high spectral resolution for the study of the inner nucleus has become available
in the X-ray band thanks to the Chandra and XMM-Newton grating spectrometers.
The first Chandra spectrum of a Seyfert 1 galaxy (\object{NGC~5548}, Kaastra et
al.  \cite{kaastra2000}) showed a wealth of discrete absorption lines.

Here we reanalyze the Chandra data of \object{NGC~5548} using significantly
improved wavelength and effective area calibration for the LETGS instrument.  We
also extend the analysis to the full wavelength range and analyze the data with
a global spectral fitting method.

The layout of the paper is as follows.  We first discuss the data analysis in
Sect.~\ref{sect:data}.  In Sect.~\ref{sect:cont} we characterize the global
continuum spectrum, and in Sect.~\ref{sect:line} we analyze individual
absorption and emission line features.  In Sect.~\ref{sect:slab} we present the
results of our global fit as well as the subsequent analysis of column densities
in terms of photoionization models.  We discuss our results in
Sect.~\ref{sect:discussion}.

\section{Data analysis\label{sect:data}}

\subsection{LETGS data}

The present Chandra observations of NGC~5548 were obtained on December 11/12,
1999, with an effective exposure time of 86400~s.  We used the High Resolution
Camera (HRC-S) in combination with the Low Energy Transmission Grating (LETG).
The spectral resolution of the instrument is about $40$~m\AA\ and almost
constant over the entire wavelength range ($1.5$--$180$~\AA).  Event selection
and background subtraction were done using the same standard processing as used
for the first-light observation of Capella (Brinkman et al.  \cite{brinkman}).

\begin{figure}
\resizebox{\hsize}{!}{\includegraphics[angle=-90]{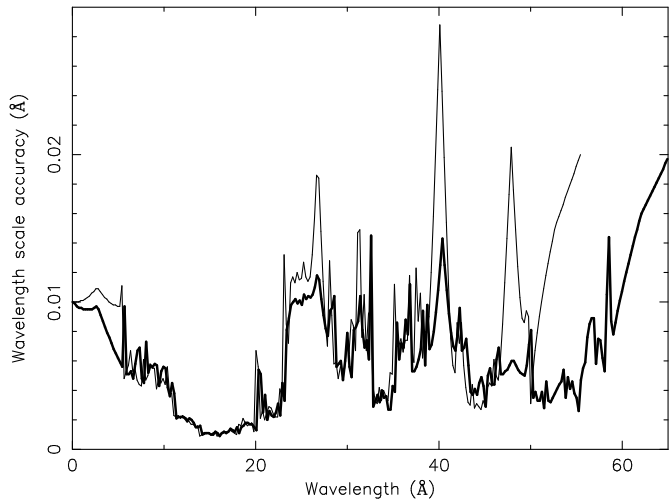}}
\caption{Accuracy of the wavelength scale for the central plate of the HRC-S
detector, for $-$1 order (thin line) and $+$1 order (thick line).}
\label{fig:wavacc}
\end{figure}

In our first analysis (Kaastra et al.  \cite{kaastra2000}) the wavelength scale
was accurate to within $15$~m\AA.  We were able to improve this accuracy by an
order of magnitude in several important regions of the spectrum by using the
strongest spectral lines of a long Capella observation.  Due to the dithering of
the instrument, these monochromatic lines sweep over the detector surface,
thereby allowing us to map the detector non-linearities.  Full details will be
described elsewhere (Van der Meer et al.  \cite{vandermeer}).  In
Fig.~\ref{fig:wavacc} we show the estimated accuracy of our wavelength scale.
Fortunately, the strong spectral line features of NGC~5548 correspond to
wavelengths for which Capella has strong emission lines.  This allows for an
accurate wavelength scale in the region with spectral features.

The line spread function of the instrument is well approximated by the square of
a Lorentzian, with a wavelength dependent width.  This shape and the relevant
parameters were derived from the ground calibration and verified in-flight using
Capella data.

In our first analysis the effective area calibration of the LETGS had not yet
been finished.  Our efficiency estimates were based upon pre-flight estimates
for wavelengths below $60$~\AA\ and on in-flight calibration using data from
Sirius~B for the longer wavelengths.  Since then a better calibration has been
obtained using LETGS observations of the white dwarf HZ\,43 and simultaneous
XMM/Chandra observations of the BL Lac object PKS\,2155-304.  The absolute
effective area is accurate to about 10~\%.  The relative effective area (on the
\AA\ scale) is accurate to within a few percent, without significant small scale
variations.

The relative strength of the higher spectral orders were derived from a
combination of ground calibration and in-flight calibration of Capella.  In our
case the combined continuum produced by the higher orders becomes comparable to
the first order continuum at wavelengths above $83$~\AA.  At these wavelengths
our signal is already very weak due to the Galactic absorption and high
background.  At $83$~\AA\ the subtracted background is ten times the first order
signal, but since the background as a function of wavelength is rather smooth,
its main effect is merely an increased noise level rather than a source of
systematic uncertainty.  Below $64$~\AA\ the subtracted background is smaller
than the measured continuum.

More details about the calibration are given in Van der Meer et al.
(\cite{vandermeer}).

A response matrix was generated including higher spectral orders up to the
$\pm$10th order for the further analysis.  All subsequent spectral fitting was
done using the SPEX package (Kaastra et al.  \cite{kaastra96}).
All errors quoted in this paper are 1~$\sigma$ errors.

\subsection{HETGS data}

NGC~5548 was observed with the HETGS instrument of Chandra two months after our
LETGS observation (Yaqoob et al.  \cite{yaqoob}; George \cite{george2001}).  We
have extracted the spectrum from the Chandra archive and processed it using the
standard CIAO pipeline.  The exposure time of this HETGS observation was similar
to the exposure time of the LETGS.  However the source was two times fainter
during the HETGS observation.  This combined with the smaller effective area of
the HETGS at longer wavelengths makes a comparison with our LETGS observations
difficult.  We have folded our best-fit warm absorber model (see
Sect.~\ref{sect:line}) through the HETGS response matrix and allowed only the
underlying continuum parameters to vary (power law and modified blackbody).  We
find no evidence for large differences between our model and the HETGS data.
Therefore we do not discuss the HETGS data in full detail.  We only use it where
it can supplement the LETGS data, in particular for the profiles of the short
wavelength lines and some emission features.

\subsection{BeppoSAX data}

BeppoSAX observed NGC~5548 simultaneously with the LETGS.  The BeppoSAX data are
especially useful for monitoring the continuum variability, the Fe-K complex
and the high energy cut-off.  Since the main aim of our paper is high resolution
X-ray spectroscopy we do not discuss these BeppoSAX data in the present paper.

\section{Continuum spectrum\label{sect:cont}}

\begin{figure}
\resizebox{\hsize}{!}{\includegraphics[angle=-90]{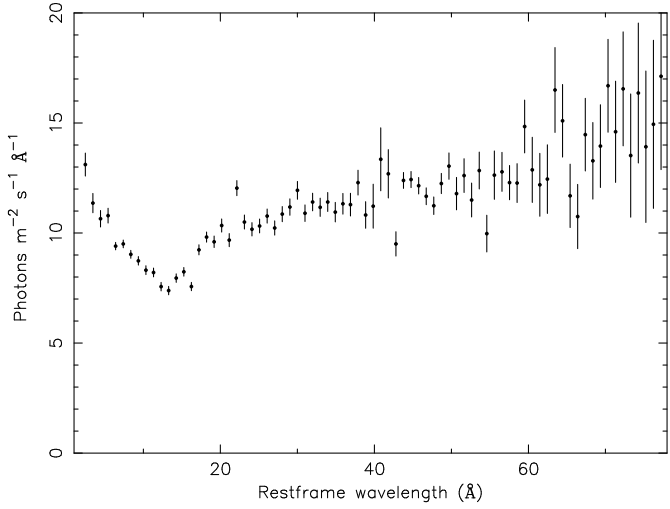}}
\caption{Fluxed, higher order and background subtracted
LETGS spectrum of NGC~5548, binned to 1~\AA\ bins
and corrected for cosmological redshift and Galactic absorption.}
\label{fig:restflux}
\end{figure}

\subsection{Fluxed spectrum}

As a first step we present in Fig.~\ref{fig:restflux} a fluxed spectrum of
NGC~5548.  Higher spectral orders have been subtracted using a bootstrap method
and the spectrum has been corrected for cosmological redshift ($z=0.01676$,
Crenshaw et al.  \cite{crenshaw}) and Galactic absorption (column density
$1.65\times 10^{24}$~m$^{-2}$).

\subsection{Fitting the continuum}

The continuum spectrum cannot be fit satisfactorily with a simple power law
model.  The broad dip in the spectrum around $15$~\AA\ clearly contradicts such
a picture.  This dip has been attributed in the past to continuum absorption
edges mainly from \ion{O}{vii} and \ion{O}{viii}.  However, the analysis of the
absorption lines as presented by Kaastra et al.  (\cite{kaastra2000}) and
refined in the present work shows that the continuum edges are much weaker.  The
dip must be attributed therefore to the high-energy cut-off of the soft excess
component.  We therefore add a modified blackbody component (MBB) to the
spectrum, in addition to the power law component, similar to the model used by
Kaastra \& Barr (\cite{kaastra89}) in their analysis of the EXOSAT data of
NGC~5548.

For a proper analysis of the continuum spectrum it is important to account for
the strongest absorption lines.  Therefore in our initial fit we excluded the
spectral regions where the absorption lines are.  If these lines are not taken
into account in the modeling (i.e., not excluded), the normalization of the
continuum is 5$-$10~\% too small, and the power law photon index is 0.05
smaller.  The lines that were excluded are given in Table~\ref{tab:lines}.  The
actual fit was done using the full resolution count spectrum.

\begin{figure}
\resizebox{\hsize}{!}{\includegraphics[angle=-90]{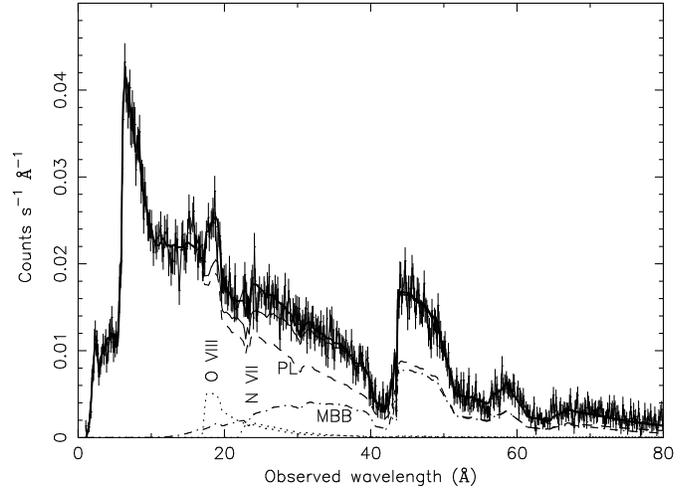}}
\caption{Chandra LETGS spectrum of NGC~5548, with absorption lines excluded
and in this plot binned by a factor of 4 in order to have a
better display of the continuum shape.
The contribution of the power law (PL) and modified blackbody (MBB)
components are indicated by the dashed and dash-dotted lines. The sum of these
two components is the thin solid line. The contribution
from the broadened \ion{O}{viii} and \ion{N}{vii} lines (dotted lines)
are indicated separately, and the total continuum is indicated by the thicker
solid line.}
\label{fig:cont}
\end{figure}

\begin{table}[!h]
\caption{Fits to the continuum only, excluding lines}
\label{tab:contfit}
\centerline{
\begin{tabular}{|lr|}
\hline
Parameter & value \\
\hline
$\chi^2$ & 1986 \\
degrees of freedom & 1834 \\
Power law photon index $\Gamma$ & 1.731$\pm$0.028 \\
Power law normalization$^{\mathrm{a}}$ & 106.0$\pm$2.0 \\
2$-$10~keV Flux (W\,m$^{-2}$) & $4.0\times 10^{-14}$ \\
MBB Temperature (keV) & 0.095$\pm$0.004 \\
MBB Normalization$^{\mathrm b}$  &
     $(2.3\pm 0.4)\times 10^{33}$\\
C VI Ly$\alpha^{\mathrm{c}}$ 33.736~\AA\ & 0($<$3) \\
N VII Ly$\alpha^{\mathrm{c}}$ 24.781~\AA\ & 12$\pm$5 \\
O VIII Ly$\alpha^{\mathrm{c}}$ 18.969~\AA\ & 19(-6,+10) \\
C VI Ly$\alpha^{\mathrm{d}}$  33.736~\AA\ & $<$0.3  \\
N VII Ly$\alpha^{\mathrm{d}}$ 24.781~\AA\ & 1.3   \\
O VIII Ly$\alpha^{\mathrm{d}}$ 18.969~\AA\ & 2.3  \\
Inclination $i$ (degrees) & 46(-3,+8) \\
Inner radius $r_1^{\mathrm{e}}$ & 1.6(-0.4,+1.0) \\
Outer radius $r_2^{\mathrm{e}}$ & 40(-30,+$\infty$) \\
Scale height $h^{\mathrm{d}}$ & 0.0 (-0.0,+1.2) \\
Emissivity index $q$ & 3.9$\pm$0.6 \\
\hline\noalign{\smallskip}
\end{tabular}
}
\begin{list}{}{}
\item[$^{\mathrm{a}}$] in photons\,m$^{-2}$\,s$^{-1}$\,keV$^{-1}$ at 1~keV
\item[$^{\mathrm{b}}$] Emitting area $A$ times square root of electron
density $n_{\mathrm e}$ in m$^{0.5}$
\item[$^{\mathrm{c}}$] Observed line intensities in units of 
photons\,m$^{-2}$\,s$^{-1}$ corrected for Galactic absorption
\item[$^{\mathrm{d}}$] Equivalent width in \AA\
\item[$^{\mathrm{e}}$] in units of $GM/c^2$
\end{list}
\end{table}

The best fit to this continuum spectrum (including weak broad emission line
features as discussed below) has a $\chi^2$ of 1986 for 1834 degrees of freedom
(Table~\ref{tab:contfit}).  The spectrum is shown in Fig.~\ref{fig:cont}.  For
the temperature of the modified blackbody component we obtained a value of
$0.095\pm0.004$~keV, the same value as derived for the EXOSAT data of July 1984
as presented by Kaastra \& Barr (\cite{kaastra89}).  Also the power law
normalization and photon index are the same as in July 1984.  However the
derived normalization of the modified blackbody component is half the value
measured by EXOSAT.

\begin{figure}
\resizebox{\hsize}{!}{\includegraphics[angle=-90]{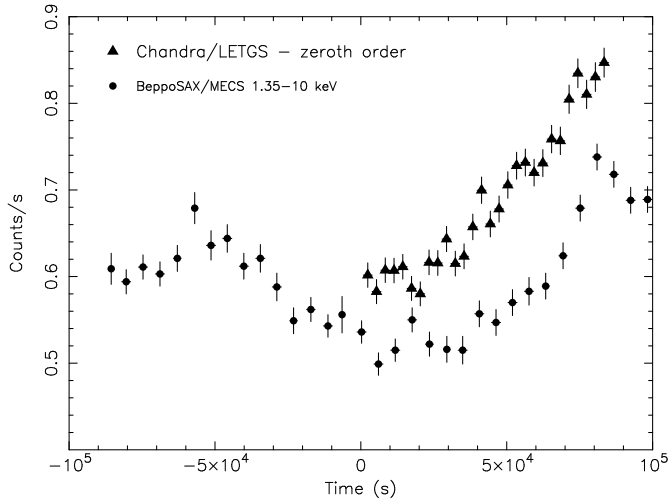}}
\caption{Light curve of NGC~5548 for the zeroth spectral order of the
LETGS and the BeppoSAX MECS data. Bin size of the LETGS data is 3000~s,
the MECS data have been averaged per satellite orbit. Zero point is
the start time of the LETGS data.}
\label{fig:lightcurve}
\end{figure}

NGC~5548 was slightly variable during the LETGS observation.  The source showed
a smooth $\sim$50~\% increase from start to end (see Fig.~\ref{fig:lightcurve}).
From an analysis of the hardness ratio in the LETGS data we find that the
spectrum softens when the source becomes brighter.  The photon index increases
from typically 1.65 at the start of the observation to 1.80 at the end.  The
same tendency is confirmed by simultaneous BeppoSAX observations.  These
BeppoSAX observations show that the source was at a minimum when the LETGS
observation started.  Since the changes in the shape of the continuum are not
very large, even though the flux increased by $\sim$50~\%, we do not investigate
the time variability any further here.  The statistical quality of the spectrum
does not allow us to check for time variability of narrow spectral features.

Finally, we note that during the HETGS observation of NGC~5548 the source showed
no significant variability at all.

\subsection{Broad emission lines}

We also tested for the possible presence of relativistically broadened Lyman
alpha lines from oxygen, nitrogen and carbon.  Such lines were recently
discovered in the XMM-Newton spectra of two narrow line Seyfert 1 galaxies
(Branduardi-Raymont et al.  \cite{branduardi}).  We added three narrow emission
lines at the rest wavelengths of the \ion{C}{vi}, \ion{N}{vii} and \ion{O}{viii}
Ly$\alpha$ lines to the model.  These lines were then convolved with the
relativistic disk line profile of Laor (\cite{laor}).  Parameters of this disk
model are the disk inner and outer radii $r_1$ and $r_2$, its inclination $i$
and the emissivity law that we take is proportional to $(r^2+h^2)^{-q/2}$ with
$q$ the emissivity power index at large radii $r$ and $h$ a scale height.

Our fit improved ($\chi^2$ decreases from 2083 to 1986 with 8 degrees of freedom
less) by including the oxygen and nitrogen lines.  We found no evidence for
significant carbon emission with the same profile (but see below).  The best fit
parameters for the continuum and broad lines are shown in
Table~\ref{tab:contfit}.

Inspecting the fit residuals of a global fit to the spectrum including the
narrow absorption lines (Sect.~\ref{sect:fitslab}) we find evidence for excess
emission around the \ion{C}{vi} Ly$\alpha$ line at $33.736$~\AA\ (see
fig.~\ref{fig:sfit06}).  This broad feature can be approximated in the rest
frame of NGC~5548 with a Gaussian profile centered at $33.84\pm 0.26$~\AA, a
width ($\sigma$) of $0.51\pm 0.16$~\AA, corresponding to $\sigma_{\mathrm v} =
4500\pm 1400$~km/s, and a flux of
5.6$^{+3.5}_{-1.7}$~photons\,m$^{-2}$\,s$^{-1}$, corrected for Galactic
absorption.  The width is significantly smaller than the width of a relativistic
line based upon the best fit model for the \ion{O}{viii} and \ion{N}{vii}
Ly$\alpha$ lines.

\section{Narrow absorption lines\label{sect:line}}

\subsection{Line identifications}

As a next step we study the narrow absorption lines.  We determined the line
centroids and equivalent widths of the most prominent features actually seen in
the spectrum as well as for some important lines that are expected to be present
but are weak.  Therefore some lines have a poorly determined equivalent width.
These lines are listed with their Doppler velocity (Sect.~\ref{sect:lineshifts})
and equivalent width in Table~\ref{tab:lines}.

\begin{table}[!h]
\caption{Absorption lines in NGC~5548. Lines significantly blended
by lines from other ions are indicated by a *; 
the equivalent widths in this case are not corrected for blending.
Unresolved line blends of
the same ion are indicated by "blend". }
\label{tab:lines}
\centerline{
\begin{tabular}{|lrrl|}
\hline
Rest frame  & velocity   & equiv. & identification \\
Wavelength & (100 km/s) & width &  \\
(\AA)      &            & (m\AA)   &  \\
\hline
40.268 & -3.7$\pm$1.6 & 136$\pm$41 & C V 1s$^2$ - 1s2p $^1$P$_1$ \\
34.973 & -2.9$\pm$3.3 &  26$\pm$22 & C V 1s$^2$ - 1s3p $^1$P$_1$ \\
\hline
33.736 & -4.3$\pm$1.2 &  90$\pm$30 & C VI 1s - 2p (Ly$\alpha$) \\
28.466 & -5.1$\pm$1.3 &  28$\pm$11 & C VI 1s - 3p (Ly$\beta$) \\
26.990 & -3.1$\pm$2.7 &  15$\pm$11 & C VI 1s - 4p (Ly$\gamma$) \\
\hline
28.787 & -3.3$\pm$3.6 &  44$\pm$24 & N VI 1s$^2$ - 1s2p $^1$P$_1$ \\
24.898 &              &   8$\pm$12 & N VI 1s$^2$ - 1s3p $^1$P$_1$ \\
\hline
24.781 & -2.9$\pm$1.9 &  66$\pm$12 & N VII 1s - 2p (Ly$\alpha$) \\
20.910 &              &   4$\pm$11 & N VII 1s - 2p (Ly$\beta$) \\
\hline
22.05  & -7.1$\pm$2.0 &  17$\pm$11 & O VI 1s$^2$2s - 1s2p($^1$P)2s \\
21.87  &              &  10$\pm$12 & O VI 1s$^2$2s - 1s2p($^3$P)2s \\
\hline
21.602 & -7.8$\pm$2.1 &  60$\pm$12 & O VII 1s$^2$ - 1s2p $^1$P$_1$ \\
18.627 & -4.4$\pm$1.0 &  34$\pm$6  & O VII 1s$^2$ - 1s3p $^1$P$_1$ \\
17.768 & -5.6$\pm$2.7 &  19$\pm$8  & O VII 1s$^2$ - 1s4p $^1$P$_1$ \\
17.396 & -3.5$\pm$3.5 &  21$\pm$9  & O VII 1s$^2$ - 1s5p $^1$P$_1$ \\
17.200 &              &  12$\pm$9  & O VII 1s$^2$ - 1s6p $^1$P$_1$ \\
\hline
18.969 & -5.4$\pm$1.0 &  60$\pm$12 & O VIII 1s - 2p (Ly$\alpha$) \\
16.006 & -0.4$\pm$1.1 &  37$\pm$7  & O VIII 1s - 3p (Ly$\beta$) \\
15.176 & +3.8$\pm$2.0 &  25$\pm$8  & O VIII 1s - 4p (Ly$\gamma$) * \\
14.821 &              &   5$\pm$9  & O VIII 1s - 5p (Ly$\delta$)  \\
\hline
13.447 & -2.2$\pm$1.6 &  38$\pm$7  & Ne IX 1s$^2$ - 1s2p $^1$P$_1$ \\
11.544 &              &   0$\pm$10 & Ne IX 1s$^2$ - 1s3p $^1$P$_1$ *\\
11.001 & -6.0$\pm$2.7 &  21$\pm$8  & Ne IX 1s$^2$ - 1s4p $^1$P$_1$ *\\
10.765 & -2.0$\pm$3.3 &  16$\pm$8  & Ne IX 1s$^2$ - 1s5p $^1$P$_1$ *\\
\hline
12.134 & -8.1$\pm$1.7 &   38$\pm$7 & Ne X   1s - 2p (Ly$\alpha$) *\\
10.238 &              &   -4$\pm$10& Ne X   1s - 2p (Ly$\beta$)  \\
\hline
11.003 & -6.5$\pm$2.7 &  21$\pm$8  & Na X  1s$^2$ - 1s2p $^1$P$_1$ * \\
10.025 &              &  22$\pm$8  & Na XI  1s - 2p (Ly$\alpha$) \\
\hline
 9.169 &              &  12$\pm$8  & Mg XI 1s$^2$ - 1s2p $^1$P$_1$ \\
 8.421 &              &  17$\pm$7  & Mg XII 1s - 2p (Ly$\alpha$) \\
\hline
61.050 & -4.1$\pm$2.4 & 129$\pm$74 & Si VIII 2p - 3d \\
43.763 & -3.0$\pm$1.9 &  35$\pm$17 & Si XI 2s$^2$ - 2s3p \\
 6.648 &              &  12$\pm$6  & Si XIII 1s$^2$ - 1s2p $^1$P$_1$ \\
 6.182 &              &  12$\pm$6  & Si XIV 1s - 2p (Ly$\alpha$) \\
\hline
42.517 & -8.5$\pm$2.3 & 161$\pm$50 & S X 2p - 3d \\
36.398 & -3.5$\pm$3.2 &  37$\pm$24 & S XII 2p - 3d \\
\hline
58.963 &              &  49$\pm$54 & Fe XIV 3p-4d \\
52.911 & -8.5$\pm$1.7 & 102$\pm$54 & Fe XV 3s$^2$ - 3s4p \\
38.956 & -8.9$\pm$3.8 &  37$\pm$26 & Fe XV 3s$^2$ - 3s5p \\
50.350 & +1.7$\pm$2.4 &  98$\pm$36 & Fe XVI 3s-4p * \\
36.749 & -6.2$\pm$2.5 &  33$\pm$24 & Fe XVI 3s-5p \\
15.014 & -3.8$\pm$1.8 &  23$\pm$8  & Fe XVII 2p-3d \\
12.123 & -5.4$\pm$1.7 &  26$\pm$9  & Fe XVII 2p-4d * \\
11.251 &              &  10$\pm$10 & Fe XVII 2p-5d \\
10.771 & -3.6$\pm$3.3 &  16$\pm$8  & Fe XVII 2p-6d \\
13.826 & -5.0$\pm$2.2 &  22$\pm$8  & Fe XVII 2p-3p \\
14.207 & -5.7$\pm$2.7 &  15$\pm$8  & Fe XVIII 2p-3d blend \\
13.522 & -3.1$\pm$2.4 &  27$\pm$8  & Fe XIX 2p-3d blend \\
12.837 &              &  35$\pm$16 & Fe XX 2p-3d blend \\
12.286 & +2.0$\pm$3.4 &  17$\pm$8  & Fe XXI 2p-3d * \\
11.706 & +2.8$\pm$3.6 &  13$\pm$9  & Fe XXII 2p-3d \\
\hline\noalign{\smallskip}
\end{tabular}
}
\end{table}

We included the \ion{O}{vi} KLL 1s2s2p resonances at $21.87$ and $22.05$~\AA\
that were recently calculated by Pradhan (\cite{pradhan}).  The slight redshift
of the \ion{O}{viii} Ly$\gamma$ line at $15.176$~\AA\ is possibly due to
contamination by \ion{Fe}{xvii} at $15.265$~\AA.

There is a problem with the \ion{Ne}{ix} 1s$^2$--1s$n$p resonance lines.  The
$n$=2 line at $13.447$~\AA\ is clearly detected.  This line may have some
contamination from the \ion{Fe}{xix} line at $13.453$~\AA.  However, there is no
significant detection of the $n$=3 line at 11.544~\AA\ (equivalent width $0\pm
10$~m\AA).  The line observed at $10.979$~\AA\ can be either \ion{Na}{x}
1s$^2$--1s2p at $11.003$~\AA\ or \ion{Ne}{ix} 1s$^2$--1s4p at $11.001$~\AA.
However given the non-detection of the \ion{Ne}{ix} $n$=3 line we prefer the
\ion{Na}{x} identification.  Similarly, we prefer to identify the line at
$10.758$~\AA\ with the \ion{Fe}{xvii} line at $10.771$~\AA\ instead of the
\ion{Ne}{ix} 1s$^2$--1s5p line at $10.765$~\AA.

\subsection{Line shifts\label{sect:lineshifts}}

The improved wavelength calibration allows us to obtain more accurate outflow
velocities.  The systematic wavelength uncertainty in general is smaller than
$0.01$~\AA, but several lines have systematic uncertainties of only
$0.001$--$0.002$~\AA\ (see Fig.~\ref{fig:wavacc}).  Only in a few cases the
systematic uncertainty reaches $0.02$~\AA, in particular near $60$~\AA\ where
Capella has no or only weak lines.  For the majority of the lines, the
statistical uncertainty of the line centroid dominates the systematic wavelength
uncertainty.  For consistency we have included the systematic errors in the
entries of Table~\ref{tab:lines}.

We express the wavelength shifts as Doppler velocities with respect to the
(optical narrow line region) redshift of 0.01676.  A negative sign indicates
blue shift (outflow).  For the weaker lines the Doppler velocity is not given
since it is poorly constrained.

\begin{figure}
\resizebox{\hsize}{!}{\includegraphics[angle=-90]{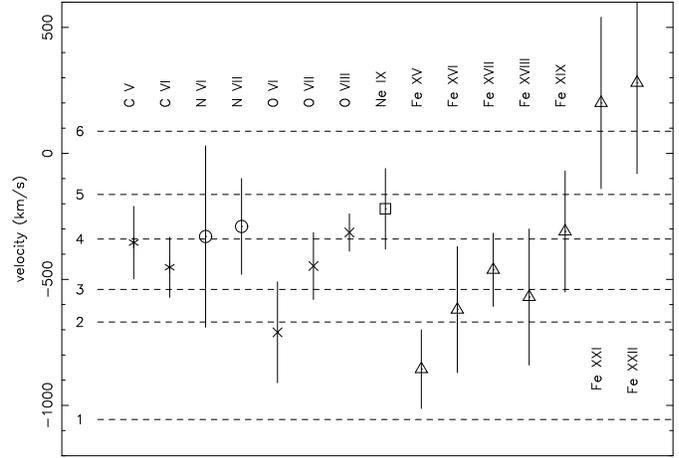}}
\caption{Average velocity with respect to the AGN rest frame for the X-ray
absorption lines of different ions. The components $1-6$ indicate the velocity
components identified in the UV by Crenshaw \& Kraemer (\cite{crenshawk}).}
\label{fig:velion}
\end{figure}

In Fig.~\ref{fig:velion} we plot the average velocity for the observed ions,
including for each ion only those lines that are not suspected to be blended
with other lines.  For \ion{O}{vii} we excluded the 1s$^2$--1s2p resonance line,
since the average wavelength of this absorption line may be contaminated by its
redshifted emission component.  We also plot the velocities of the 6 velocity
components identified in UV absorption lines by Crenshaw \& Kraemer
(\cite{crenshawk}).  The component velocities are:  $-1056$, $-669$, $-540$,
$-340$, $-163$ and $+88$~km/s.

Most of the ions observed in the X-rays have velocities consistent with the
velocity of component 4 ($-340$~km/s).  However a velocity distribution between
components 2$-$5 is also possible.  For the iron ions we see a clear tendency
for less ionized ions to have larger outflow velocities than the more highly
ionized ions.  The same tendency is visible for the oxygen ions.  A similar
velocity pattern was also found in the XMM-Newton spectrum of the quasar
IRAS~13349+2438 by Sako et al.  (\cite{sako2001}).

\subsection{Line profiles and broadening\label{sect:linebroad}}

As was shown by Kaastra et al.  (\cite{kaastra2000}) the measured Doppler width
of the absorption lines was larger than the width derived from the intensity
ratio of different lines from the same ion.  This provides evidence that the
lines are actually blends composed of different velocity components, with
varying relative strength for each ion or group of ions.  Our LETGS data have
insufficient spectral resolution to resolve the blends fully, although we see
evidence for broadening, differences in the centroids of the blends etc.

\begin{figure}
\resizebox{\hsize}{!}{\includegraphics[angle=-90]{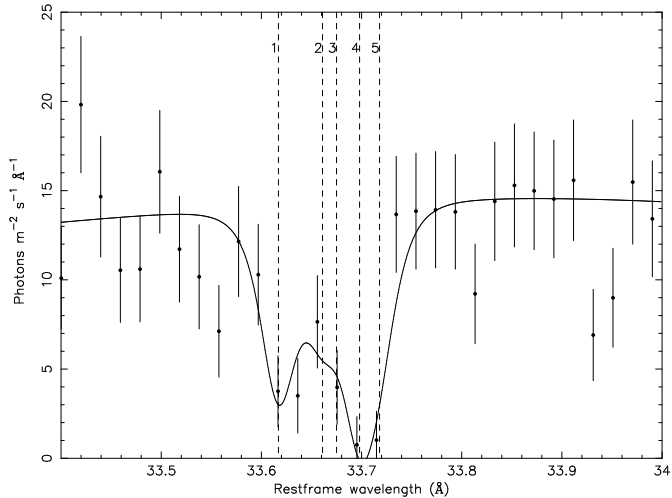}}
\caption{The \ion{C}{vi} Ly$\alpha$ line fitted using a sum of narrow
line components with the velocities fixed at the values for the velocity
components 1--5 identified in the UV by Crenshaw \& Kraemer (\cite{crenshawk}).}
\label{fig:c6line}
\end{figure}

This is illustrated in Fig.~\ref{fig:c6line}, where we fitted the \ion{C}{vi}
Ly$\alpha$ line using a sum of narrow line components with the velocities fixed
at the values for the velocity components 1--5 identified in the UV by Crenshaw
\& Kraemer (\cite{crenshawk}).  The \ion{C}{vi} Ly$\alpha$ line is the best case
to study here, since it is strong, lies in a region with good statistics and
relatively high spectral resolution of the LETGS.  It is also well isolated (not
blended).  In Table~\ref{tab:c6line} we show the best-fit equivalent widths for
the individual components.  The centroids of these 5 components span only two
times the instrumental FWHM.  As a result there is a strong correlations between
these parameters and the equivalent widths are poorly constrained.

\begin{table}[!h]
\caption{\ion{C}{vi} Ly$\alpha$ line.}
\label{tab:c6line}
\centerline{
\begin{tabular}{|lrr|}
\hline
Component & velocity & equivalent width \\
 & km/s & m\AA \\
\hline
1 & -1056 & $37\pm 10$ \\
2 & -669  & $19\pm 26$ \\
3 & -540  & $0\pm 31$ \\
4 & -340  & $37\pm 21$\\
5 & -163  & $20\pm 15$\\
Total & & $113\pm49$ \\
\hline\noalign{\smallskip}
\end{tabular}
}
\end{table}

\begin{figure}
\resizebox{\hsize}{!}{\includegraphics[angle=-90]{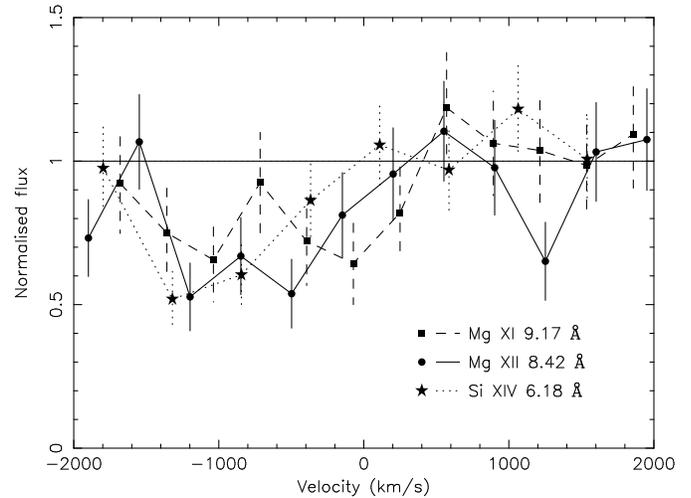}}
\caption{Line profiles of the strongest lines in the MEG spectrum
of NGC~5548. Fluxes have been normalized to 1 for the continuum.
Bin size is $0.01$~\AA.}
\label{fig:megprof}
\end{figure}

For comparison we also plot the velocity profiles of the strongest short
wavelength lines as measured with the Medium Energy Grating (MEG) of the HETGS
observation (Fig.~\ref{fig:megprof}).  The equivalent widths of these absorption
lines are within their error bars consistent with the values obtained from the
LETGS in Table~\ref{tab:lines}.  The Mg lines have a significant contribution
from the lower velocity gas components 2--5 (few hundred km/s outflow), but in
addition the Mg lines and in particular the \ion{Si}{xiv} Ly$\alpha$ line at
$6.18$~\AA\ have a strong component around $-1000$~km/s, consistent with
component 1. Apparently \ion{Si}{xiv} does not follow the general trend
seen in the iron ions (Fig.~\ref{fig:velion}).

Given this complex line structure, we may question if an analysis using a single
line component in order to determine the column densities is sufficiently
accurate.

In the UV, with 100 times better spectral resolution, the line blends are
clearly resolved into 6 velocity components (Crenshaw \& Kraemer
\cite{crenshawk}) as we noted before.  However, the individual components show
further substructure.  This is particularly evident from the profile of the
dominant UV component (4) in \ion{C}{iv} and \ion{N}{v}.  In their figure 1 this
component has a rather "square", non-Gaussian appearance.  Mathur et al.
(\cite{mathur99}) split this component into two parts.

An important consideration is the number of velocity components taken into
account, especially for saturated absorption lines.  Consider for example a set
of $n$ equal absorbers, each having a large column density $N$ that produces
sufficient line opacity for the line core of each component to be saturated.  If
each absorber has the same velocity, the total equivalent width
$W_{\mathrm{tot}}$, due to the overlapping line profiles, is significantly
smaller than $n$ times the equivalent width $W_{\mathrm{ind}}$ for an individual
component.  But if the velocity separations of the components are larger than
the intrinsic (turbulent or thermal) line width $\sigma_{\mathrm v}$, then
$W_{\mathrm{tot}} = nW_{\mathrm{ind}}$.

In order to take this effect into account, we adopt the following procedure in
the column density analysis of the next section.  We split each line into 5
velocity components with equal column density for each component.  These 5
components have the same intrinsic line broadening $\sigma_{\mathrm v}$ and have
fixed velocities of $-163$, $-305$, $-375$, $-540$ and $-669$~km/s corresponding
to the UV components 5, 4a, 4b, 3 and 2 respectively, where we split the
strongest and broadest component 4 ($-340$~km/s) into two parts 4a and 4b, cf.
Mathur et al.  (\cite{mathur99}).  Of course this treatment ignores possible
differences between the UV and X-ray lines, but since for most lines we do not
have sufficient spectral resolution to resolve the profiles it is the best we
can do with the present data set.

\section{Detailed spectral fitting\label{sect:slab}}

In Sect.~\ref{sect:line} we have identified the most prominent absorption lines.
From the observed equivalent width ratio of these absorption lines it is
possible to derive the column densities of the absorbing ions (e.g.  Kaastra et
al.  (\cite{kaastra2000}).  There is good reason, however, to go one step
further.  First, the continuum opacity of the warm absorber appears to be small
but not completely negligible.  This may affect the estimate of the continuum
spectrum.  Further, some lines are significantly blended by lines from other
ions resulting in uncertainties in the equivalent widths.  Finally, some ions
produce absorption lines that are too weak to be detected individually, but
taken together these lines may represent a detectable signal.  Moreover, column
densities can also be constrained by lines that are predicted to be present but
are not detected.  Therefore we fitted the complete LETGS spectrum starting with
the continuum model of Sect.~\ref{sect:cont} with the addition of a warm
absorber model as described below.

\subsection{Slab model\label{sect:slabmodel}}

We generated a new model, named {\it slab}, for the SPEX package that treats in
a simplified manner the absorption by a thin slab composed of different ions.
The transmission $T(\lambda)$ of this slab is simply calculated as
$T(\lambda)=\exp[{-\tau_c(\lambda)-\tau_l(\lambda)}]$ with $\tau_c$ and $\tau_l$
the total continuum and line optical depth, respectively.  As long as the
thickness of the slab is not too large, this most simple approximation allows a
fast computation of the spectrum, which is desirable for spectral fitting.

The continuum opacities are taken from Verner \& Yakovlev (\cite{verner95}).
Line opacities and wavelengths for most ions are from Verner et al.
(\cite{verner96}), with the following additions.  For \ion{O}{vi} we added the
KLL resonances near $22$~\AA\ as calculated by Pradhan (\cite{pradhan}).  For
the L-shell ions of iron (\ion{Fe}{xvii} $-$ \ion{Fe}{xxiv}) we used the HULLAC
calculations provided by one of us (Liedahl).  The wavelengths of these L-shell
transitions were adjusted according to Phillips et al.  (\cite{phillips}).  Also
the L-shell ions of \ion{Si}{viii}--\ion{Si}{xii}, \ion{S}{x}--\ion{S}{xiv} and
\ion{Ar}{xv} were calculated using the HULLAC code.  In addition, we added the
2p--3d inner shell resonance lines of \ion{Fe}{vii}$-$\ion{Fe}{xv} that we have
calculated using the Cowan code (\cite{cowan}).  These data were compared with
those in Behar et al.  (\cite{behar}) and found to be in reasonable agreement.
These inner shell resonance lines occur mainly in the $16$--$17$~\AA\ band and
appear to be extremely important diagnostic tools, since the normal resonance
lines of these lowly ionized iron ions have their wavelengths mainly in the
inaccessible EUV band.  Their importance for active galactic nuclei was first
demonstrated by Sako et al.  (\cite{sako2001}).

Each absorption line is then split into different velocity components, using
\begin{equation}
\tau_l(v) = \sum\limits_{i}^{}\tau_i \exp\left[ 
-(v-v_i)^2/2\sigma_{\mathrm v}^2
   \right].
\end{equation}

In our implementation for NGC~5548, we have chosen the velocities $v_i$ of the
velocity components as we discussed at the end of Sect.~\ref{sect:linebroad}
(i.e.  five components at outflow velocities of $-163$, $-305$, $-375$, $-540$
and $-669$~km/s).  We also took $\tau_i=\tau_0 / 5$ for all these components,
with $\tau_0$ the total optical depth given by
\begin{equation}
\label{eqn:tau}
\tau_0 = 0.106 f N_{20} \lambda / \sigma_{\rm v,100}.
\end{equation}
Here $f$ is the oscillator strength, $\lambda$ the wavelength in \AA,
$\sigma_{\rm v,100}$ the velocity dispersion in units of $100$~km/s and $N_{20}$
the total column density of the ion in units of $10^{20}$~m$^{-2}$, added for
all five components.

We applied this absorber model to the continuum model described before and
fitted the entire spectrum.  In addition, we have included in the model the
strongest narrow emission components for the \ion{O}{vii} triplet and the
\ion{Ne}{ix} forbidden line as found by Kaastra et al.  (\cite{kaastra2000}).

In a more general implementation of this model (not used here for NGC~5548)
we take instead
\begin{equation}
v_i = v_0 + i\,\Delta v,
\end{equation}
\begin{equation}
\label{eqn:taui}
\tau_i = K \exp\left[ -v_i^2/2b^2 \right],
\end{equation}
where $v_0$ is the average velocity of the blend (a negative value corresponds
to a blue-shift or outflow), $\Delta v$ is the separation between the components,
and the r.m.s.  width of the blend $b$ is in general larger than the intrinsic
width $\sigma_{\mathrm v}$ of the components.  The normalization $K$ is defined
in such a way that $\sum \tau_i = \tau_0$.

\subsection{Fit results using the slab model\label{sect:fitslab}}

\begin{table}[!h]
\caption{Fits to the total spectrum}
\label{tab:totfit}
\centerline{
\begin{tabular}{|lr|}
\hline
Parameter & value \\
\hline
$\chi^2$ & 2211 \\
degrees of freedom & 1963 \\
Power law photon index $\Gamma$ & 1.848$\pm$0.021 \\
Power law normalization$^{\mathrm{a}}$ & 131.8$\pm$1.6 \\
MBB Temperature (keV) & 0.081$\pm$0.007 \\
MBB Normalization $^{\mathrm{b}}$ &
     $(2.3\pm 0.6)\times 10^{33}$\\
C VI Ly$\alpha^{\mathrm{c}}$ 33.736~\AA\ & 0($<$8) \\
N VII Ly$\alpha^{\mathrm{c}}$ 24.781~\AA\ & 11.6$\pm$4.0 \\
O VIII Ly$\alpha^{\mathrm{c}}$ 18.969~\AA\ & 6.5$\pm$2.3 \\
C VI Ly$\alpha^{\mathrm{d}}$ 33.736~\AA\ & $<$0.7 \\
N VII Ly$\alpha^{\mathrm{d}}$ 24.781~\AA\ & 1.1 \\
O VIII Ly$\alpha^{\mathrm{d}}$ 18.969~\AA\ & 0.6 \\
Inclination $i$ (degrees) & 46.0 \\
Inner radius $r_1\,^{\mathrm{e}}$ & 1.6 \\
Outer radius $r_2\,^{\mathrm{e}}$ & 40 \\
Scale height $h\,^{\mathrm{e}}$ & 0.0  \\
Emissivity index $q$ (fixed) & 3.9 \\
Narrow emission lines & see Table~\ref{tab:emlines_f} and \ref{tab:emlines_v} \\
Column densities & see Table~\ref{tab:ioncol} \\
\hline\noalign{\smallskip}
\end{tabular}
}
\begin{list}{}{}
\item[$^{\mathrm{a}}$] in photons\,m$^{-2}$\,s$^{-1}$\,keV$^{-1}$ at 1~keV.
\item[$^{\mathrm{b}}$] Emitting area times square root of electron
density in $m^{0.5}$.
\item[$^{\mathrm{c}}$] Observed line intensities in units of 
ph\,m$^{-2}$\,s$^{-1}$ corrected for Galactic absorption.
\item[$^{\mathrm{d}}$] Equivalent width in \AA.
\item[$^{\mathrm{e}}$] in units of $GM/c^2$; parameters fixed.
\end{list}
\end{table}

The results of our fit are summarized in Table~\ref{tab:totfit} and
Table~\ref{tab:ioncol}.  We have kept the shape parameters of the
relativistically broadened line profiles for Ly$\alpha$ of \ion{O}{viii},
\ion{N}{vii} and \ion{C}{vi} fixed to the values given in
Table~\ref{tab:contfit}, given their large uncertainties.  However the strength
of the lines was allowed to vary.  The intrinsic width of the absorption line
components was kept fixed to $\sigma_{\mathrm v}=32$~km/s (r.m.s.  width), in
agreement with the UV line components.  When we left it as a free parameter, the
best fit parameter was $47_{-12}^{+59}$~km/s, in fair agreement with our fixed
value.

\begin{table*}[!ht]
\caption{Measured and model column densities. Listed are
the logarithms of the column densities in units of m$^{-2}$.
The columns labeled A, B and C list as a percentage
the relative contribution of the three ionization components to
the total model column density. The last three columns give rest frame
wavelength, oscillator strength and optical depth at line center
for the strongest line of each ion in the LETGS or UV band, for
$\sigma_{\mathrm v}=32$~km/s and a single velocity component
with 20~\% of the
total (observed) ion column density.}
\label{tab:ioncol}
\centerline{
\begin{tabular}{|lrrr|r|l|rrr|}
\hline
Ion & A  & B  & C  & model & observed & $\lambda$ (\AA) & $f$ & $\tau_0$ \\
\hline
H  I    & 18& 45& 38&  19.65&  19.65 (-0.02,+0.02)&1215.670& 0.42& 15.0\\
C  IV   &  0&  1& 99&  19.02&  19.08 (-0.03,+0.03)&1548.195& 0.19&  2.3\\
C  V    &  1& 44& 55&  20.40&  20.56 (-0.38,+0.30)&  40.268& 0.65&  6.3\\
C  VI   & 43& 55&  1&  21.70&  21.32 (-0.32,+0.29)&  33.736& 0.42& 19.4\\
N  V    &  0&  7& 93&  18.93&  19.11 (-0.03,+0.03)&1238.820& 0.16&  1.7\\
N  VI   &  3& 81& 16&  20.10&  20.19 (-0.39,+0.34)&  28.787& 0.68&  2.0\\
N  VII  & 53& 47&  0&  21.25&  20.81 (-0.25,+0.29)&  24.781& 0.42&  4.4\\
O  VI   &  0& 36& 64&  19.98&  19.61 (-0.13,+0.13)&1031.926& 0.13&  3.7\\
        &   &   &   &       &  19.90 (-$\infty$,+0.63)&  22.007& 0.52&  0.3\\
O  VII  &  7& 91&  3&  21.25&  21.61 (-0.17,+0.14)&  21.602& 0.70& 40.6\\
O  VIII & 67& 33&  0&  22.17&  22.21 (-0.15,+0.11)&  18.969& 0.42& 84.8\\
Ne IX   & 25& 75&  0&  21.38&  21.66 (-0.31,+0.19)&  13.447& 0.72& 29.5\\
Ne X    & 91&  9&  0&  22.13&  21.44 (-0.32,+0.30)&  12.134& 0.42&  9.2\\
Na X    & 49& 51&  0&  20.29&  20.48 (-$\infty$,+0.64)&  11.003& 0.73&  1.6\\
Na XI   & 97&  3&  0&  21.03&  20.92 (-0.66,+0.48)&  10.025& 0.42&  2.3\\
Mg IX   &  0& 98&  1&  20.37&  20.30 (-0.44,+0.28)&  62.751& 0.54&  4.5\\
Mg X    &  7& 93&  0&  20.47&  18.15 (-$\infty$,+1.74)&  57.876& 0.22&  0.0\\
Mg XI   & 74& 26&  0&  21.29&  20.95 (-1.23,+0.70)&   9.169& 0.74&  4.0\\
Mg XII  & 99&  1&  0&  21.93&  22.08 (-0.48,+0.25)&   8.421& 0.42& 27.9\\
Si VIII &  0& 14& 86&  18.78&  19.62 (-1.02,+0.45)&  61.070& 0.62&  1.0\\
Si IX   &  0& 81& 19&  19.22&  20.26 (-0.57,+0.43)&  55.305& 0.95&  6.3\\
Si X    &  1& 98&  1&  20.19&  20.04 (-0.34,+0.23)&  47.489& 0.21&  0.7\\
Si XI   &  5& 95&  0&  20.55&  19.74 (-0.45,+0.29)&  43.762& 0.48&  0.8\\
Si XII  & 47& 53&  0&  20.52&  14.57 (-$\infty$,+5.83)&  31.012& 0.06&  0.0\\
Si XIII & 97&  3&  0&  21.54&  21.59 (-0.79,+0.54)&   6.648& 0.76& 13.0\\
Si XIV  &100&  0&  0&  21.84&  22.19 (-0.85,+0.36)&   6.182& 0.42& 26.4\\
S  X    &  1& 89& 11&  19.62&  19.95 (-0.36,+0.33)&  42.543& 0.67&  1.7\\
S  XI   &  3& 97&  0&  20.31&  20.29 (-0.60,+0.41)&  39.240& 1.05&  5.3\\
S  XII  & 13& 87&  0&  20.59&  19.99 (-0.48,+0.33)&  36.398& 0.62&  1.5\\
S  XIII & 43& 57&  0&  20.80&  13.88 (-$\infty$,+5.56)&  32.242& 0.45&  0.0\\
S  XIV  & 91&  9&  0&  21.06&  14.89 (-$\infty$,+5.06)&  23.015& 0.30&  0.0\\
S  XV   &100&  0&  0&  21.95&  22.19 (-0.65,+0.28)&   5.039& 0.77& 39.7\\
Ar XIII & 12& 88&  0&  19.57&  14.41 (-$\infty$,+4.41)&  29.320& 0.91&  0.0\\
Ar XIV  & 43& 57&  0&  19.65&  14.30 (-$\infty$,+5.18)&  27.470& 0.65&  0.0\\
Ar XV   & 84& 16&  0&  19.86&  20.00 (-$\infty$,+0.54)&  24.737& 0.39&  0.6\\
Fe VII  &  0&  0&100&  19.18&  19.37 (-$\infty$,+0.73)&  16.903& 0.18&  0.0\\
Fe VIII &  0&  0&100&  19.87&  19.10 (-$\infty$,+0.86)&  16.658& 0.36&  0.1\\
Fe IX   &  0&  0&100&  20.15&  18.48 (-$\infty$,+1.52)&  16.510& 1.41&  0.0\\
Fe X    &  0&  0&100&  20.28&  19.68 (-$\infty$,+0.43)&  16.339& 0.76&  0.4\\
Fe XI   &  0&  0&100&  20.14&  20.27 (-0.19,+0.15)&  16.132& 0.34&  0.7\\
Fe XII  &  0& 14& 86&  19.92&  19.80 (-$\infty$,+0.37)&  15.963& 0.42&  0.3\\
Fe XIII &  0& 87& 13&  20.23&  19.35 (-$\infty$,+0.53)&  15.849& 0.70&  0.2\\
Fe XIV  &  0& 98&  2&  20.31&  18.94 (-$\infty$,+0.70)&  15.579& 0.61&  0.1\\
Fe XV   &  0&100&  0&  20.57&  19.78 (-0.60,+0.38)&  15.316& 1.74&  1.1\\
Fe XVI  &  0&100&  0&  20.17&  20.47 (-0.44,+0.29)&  15.250& 0.96&  2.9\\
Fe XVII &  2& 98&  0&  20.64&  20.95 (-0.21,+0.18)&  15.014& 2.72& 24.1\\
Fe XVIII& 36& 64&  0&  20.42&  20.14 (-0.39,+0.27)&  14.204& 0.88&  1.1\\
Fe XIX  & 91&  9&  0&  20.52&  20.69 (-0.19,+0.16)&  13.521& 0.72&  3.1\\
Fe XX   & 99&  1&  0&  20.79&  20.64 (-0.25,+0.21)&  12.827& 0.47&  1.8\\
Fe XXI  &100&  0&  0&  20.57&  20.29 (-0.67,+0.43)&  12.286& 1.24&  2.0\\
Fe XXII &100&  0&  0&  20.31&  20.27 (-$\infty$,+0.45)&  11.706& 0.67&  1.0\\
Fe XXIII&100&  0&  0&  19.96&  20.10 (-$\infty$,+0.55)&  10.981& 0.41&  0.4\\
Fe XXIV &100&  0&  0&  19.23&  15.00 (-$\infty$,+4.72)&  10.620& 0.24&  0.0\\
\hline\noalign{\smallskip}
\end{tabular}
}
\end{table*}

In those cases where a zero column density is within the error bars, we put the
lower error on the logarithm of the column density formally to $-\infty$.  The
measured column densities agree in general within their error bars with the
values derived from the strongest individual lines as given by Kaastra et al.
(\cite{kaastra2000}).  For completeness, we also include the total column
densities for the co-added components 2--5 for \ion{H}{i}, \ion{C}{iv},
\ion{N}{v} and \ion{O}{vi} as derived from UV measurements by the Space
Telescope Imaging Spectrograph (STIS) as well as the Far Ultraviolet
Spectroscopic Explorer (FUSE).  We comment on these UV lines in
Sect.~\ref{sect:uvlines}.

\begin{figure}
\resizebox{\hsize}{!}{\includegraphics[angle=-90]{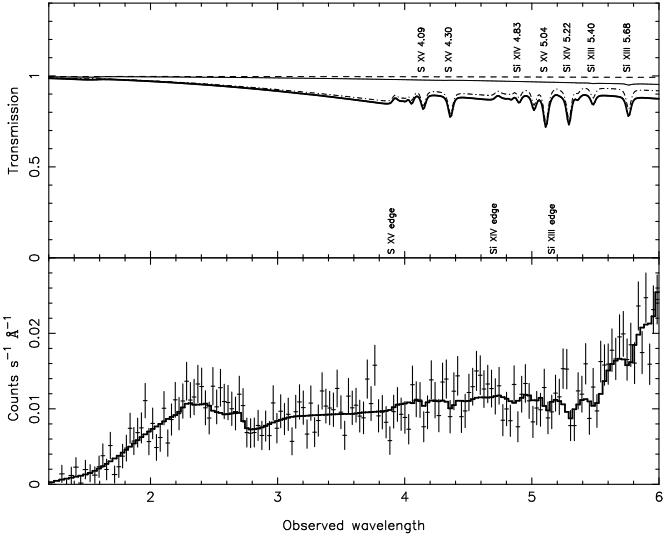}}
\caption{lower panel: observed spectrum with the best fit model.
Upper panel: transmission of the high ionization absorber
component A (dash-dotted line),
medium ionization component B (thin solid line), low ionization component
C (dashed line) and total transmission (thick solid line). Line
identifications are given with their rest frame wavelengths.
}
\label{fig:sfit01}
\end{figure}

\begin{figure}
\resizebox{\hsize}{!}{\includegraphics[angle=-90]{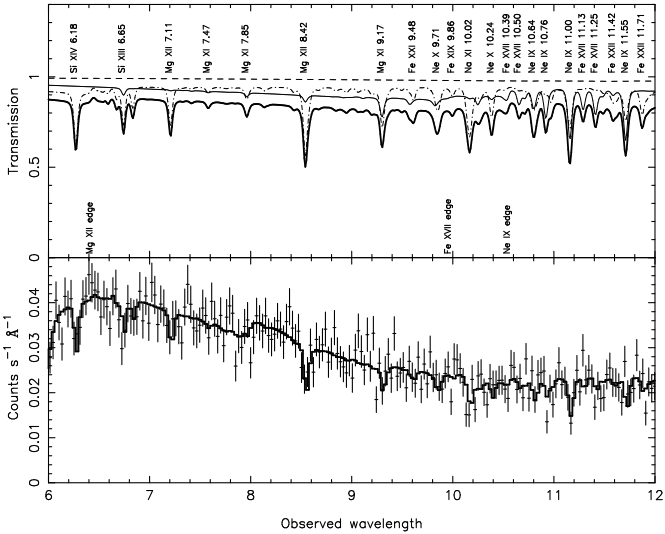}}
\caption{As Fig.~\ref{fig:sfit01}, but for $6$--$12$~\AA.}
\label{fig:sfit02}
\end{figure}

\begin{figure}
\resizebox{\hsize}{!}{\includegraphics[angle=-90]{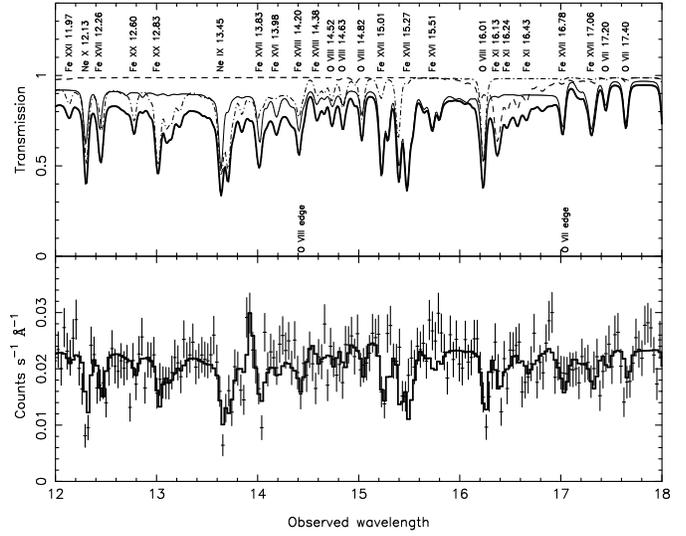}}
\caption{As Fig.~\ref{fig:sfit01}, but for $12$--$18$~\AA.}
\label{fig:sfit03}
\end{figure}

\begin{figure}
\resizebox{\hsize}{!}{\includegraphics[angle=-90]{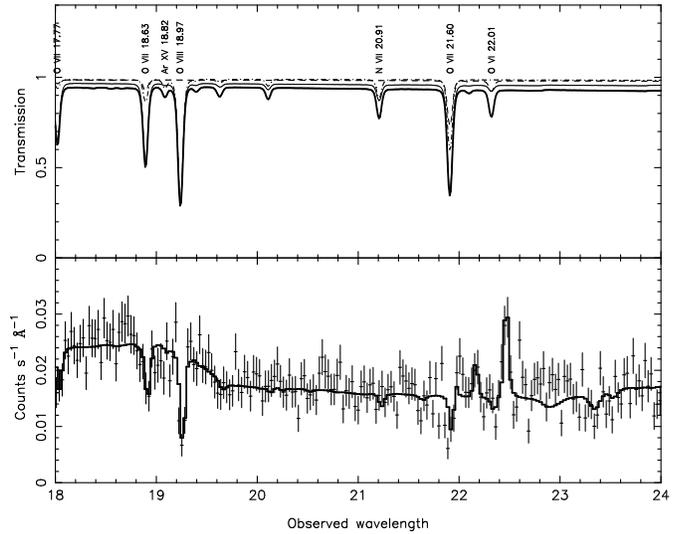}}
\caption{As Fig.~\ref{fig:sfit01}, but for $18$--$24$~\AA.}
\label{fig:sfit04}
\end{figure}

\begin{figure}
\resizebox{\hsize}{!}{\includegraphics[angle=-90]{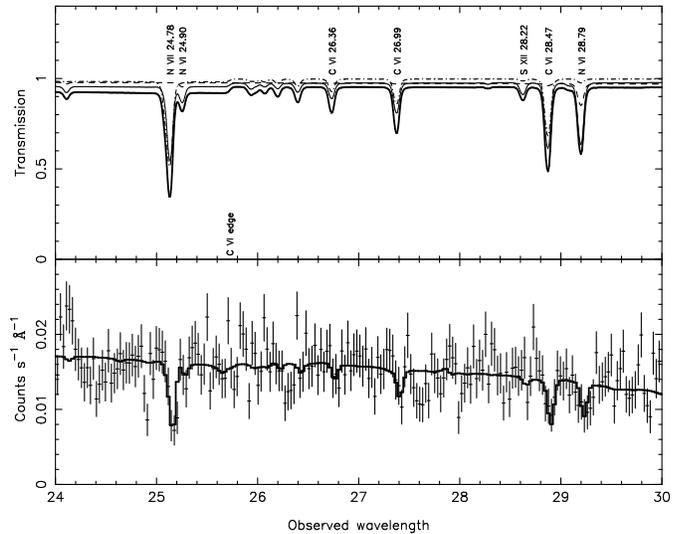}}
\caption{As Fig.~\ref{fig:sfit01}, but for $24$--$30$~\AA.}
\label{fig:sfit05}
\end{figure}

\begin{figure}
\resizebox{\hsize}{!}{\includegraphics[angle=-90]{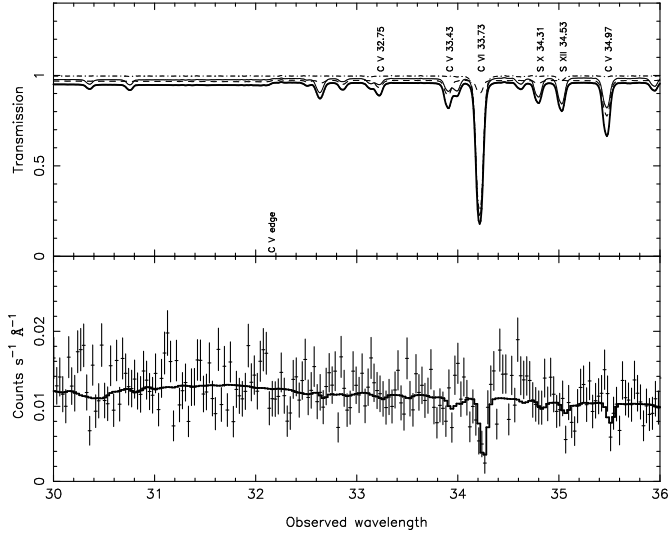}}
\caption{As Fig.~\ref{fig:sfit01}, but for $30$--$36$~\AA.}
\label{fig:sfit06}
\end{figure}

\begin{figure}
\resizebox{\hsize}{!}{\includegraphics[angle=-90]{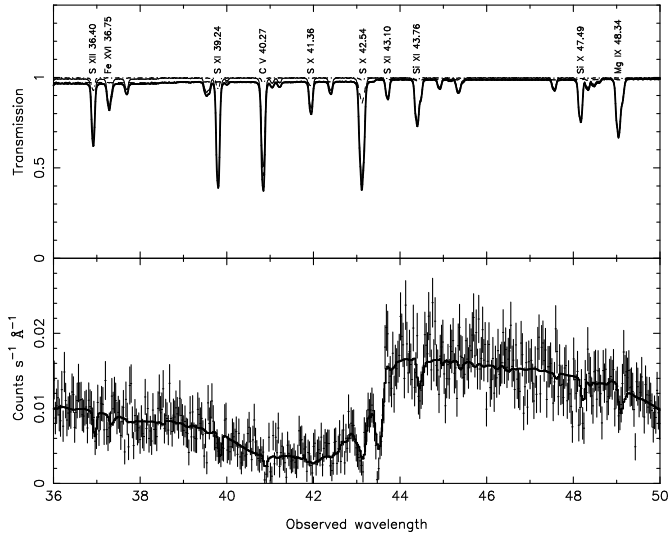}}
\caption{As Fig.~\ref{fig:sfit01}, but for $36$--$50$~\AA.}
\label{fig:sfit07}
\end{figure}

\begin{figure}
\resizebox{\hsize}{!}{\includegraphics[angle=-90]{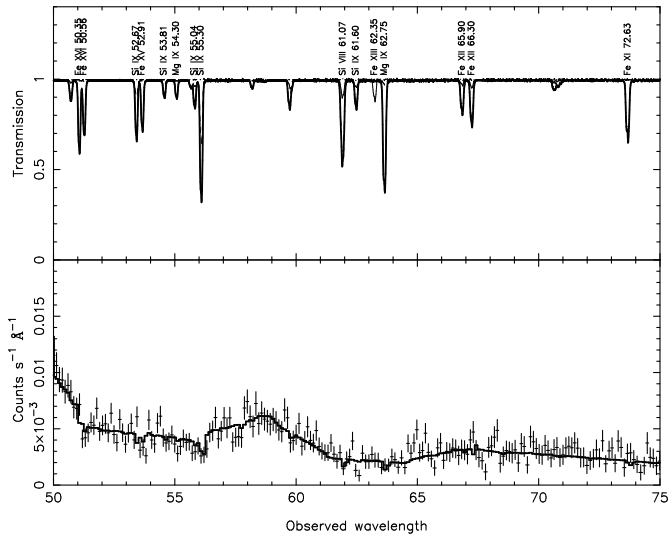}}
\caption{As Fig.~\ref{fig:sfit01}, but for $50$--$75$~\AA.}
\label{fig:sfit08}
\end{figure}

\begin{figure}
\resizebox{\hsize}{!}{\includegraphics[angle=-90]{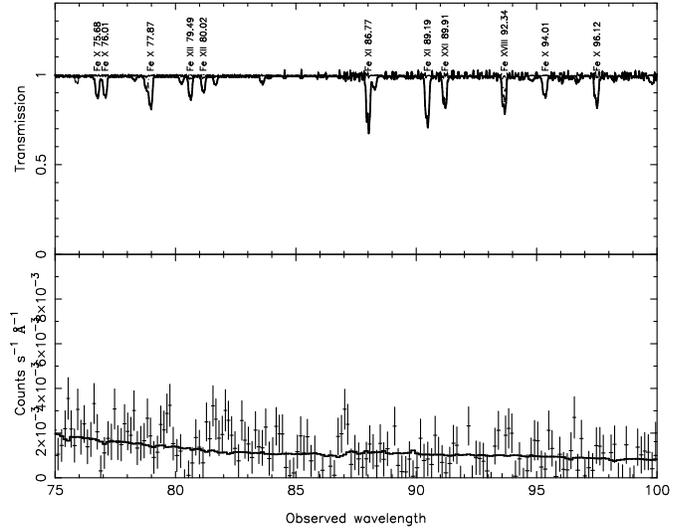}}
\caption{As Fig.~\ref{fig:sfit01}, but for $75$--$100$~\AA.}
\label{fig:sfit09}
\end{figure}

In Figs.~\ref{fig:sfit01}--\ref{fig:sfit09} we show the best fit spectrum with
the transmission of the model components included.


\subsection{Other atomic data for inner-shell transitions}

Wavelengths and oscillator strengths for the 2p--3d UTA of
\ion{Fe}{i}--\ion{Fe}{xvi} have also been calculated recently by Behar et al.
(\cite{behar}).  Their results are in good agreement with the present
calculations for these ions.  We have tested this explicitly by using the Behar
et al.  data in our spectral fits; the derived column densities are all within
the error bars consistent with the results presented in Table~\ref{tab:ioncol}.

Recently it has also become evident that inner shell transitions of other ions
such as oxygen may be important in AGN (Behar \& Netzer \cite{behar2002}).  In
our current modeling, we have taken the \ion{O}{vi} data from Pradhan
(\cite{pradhan}).  From Table~\ref{tab:ioncol} we see that the column density of
this ion is poorly constrained.  The wavelengths and oscillator strengths for
all of the inner shell 1s--$n$p ($n\le 7$) transitions in
\ion{O}{i}--\ion{O}{vi} have been recently calculated (Behar 2002, in
preparation).  Lines from these ions have recently been discovered in the RGS
spectrum of NGC~5548 (Steenbrugge et al.  \cite{steenbrugge}).  We included
those lines also in our code but did not detect significant absorption from
these ions (Table~\ref{tab:oxlines}).  Therefore we did not take these ions into
account in our further analysis, except for \ion{O}{vi} but with the value
quoted in Table~\ref{tab:ioncol}.

\begin{table}[!h]
\caption{1$\sigma$ upper limits to the oxygen column densities. Listed are
the logarithms of the column densities in units of m$^{-2}$.}
\label{tab:oxlines}
\centerline{
\begin{tabular}{|lr|}
\hline
Ion & observed \\
\hline
O I     & $<$20.13 \\
O II    & $<$20.06 \\
O III   & $<$20.68 \\
O IV    & $<$19.68 \\
O V     & $<$20.48 \\
O VI    & $<$20.44 \\
\hline\noalign{\smallskip}
\end{tabular}
}
\end{table}

\subsection{Supplementary UV data\label{sect:uvlines}}

The column densities that we derived can be compared to the predictions from
photoionization models.  Before this comparison we want to supplement the data
with UV data, which can help to constrain the lowest ionization components.

For \ion{C}{iv}, it was shown recently by Arav et al.  (\cite{arav}) that the
column density as derived by Crenshaw \& Kraemer (\cite{crenshaw}) is four times
too small.  This is due in part to the fact that apparently the narrow line
region is not covered by the warm absorber.  Also the fact that the line is
highly saturated, so that the line profile merely represents the covering factor
and not the optical depth of the absorbing gas, contributes to this.  Thus, we
adopted the column density derived by Arav et al.

For \ion{H}{i}, Crenshaw \& Kraemer obtain a logarithmic column of
18.80$\pm$0.02, based upon the Ly$\alpha$ line as observed with the STIS in
March 1998.  Recently, Brotherton et al.  (\cite{brotherton}) determined a
\ion{H}{i} column density of 19.65$\pm$0.02, based upon the Ly$\beta$ line
observed by FUSE in June 2000.  If not caused by strong variability, this last
column density would imply that the Ly$\alpha$ line, with its higher oscillator
strength, should have an even higher optical depth than the \ion{C}{iv} line.
Hence we may expect similar saturation problems for Ly$\alpha$ as for
\ion{C}{iv}.  Therefore we prefer the column density derived from the Ly$\beta$
line.

Optical depth effects are estimated to be less important for \ion{N}{v}, so for
that ion we adopt the STIS values of Crenshaw \& Kraemer (\cite{crenshaw}).

Finally, our estimate for the \ion{O}{vi} column density, based upon the present
X-ray data, is rather uncertain:  the best fit logarithmic column density is
$19.90_{-\infty}^{+0.63}$.  This value is, however, consistent with the column
density derived from the ultraviolet $1032$, $1038$~\AA\ lines:  $19.61\pm 0.13$
(Brotherton et al.  (\cite{brotherton}).  We increased the uncertainty in this
number to take into account the effects of incomplete covering of the central
source.  We used the latter more accurate value in all subsequent modeling.

\subsection{Modeling of the column density\label{sect:modcol}}

The column densities that we derived can be compared to the values predicted by
photoionization calculations.  As in Kaastra et al.  (\cite{kaastra2000}) we
have chosen to compare the column densities with calculations made with the
XSTAR code (Kallman \& Krolik \cite{kallman}).  The main difference with the
previous work is the larger amount of ions that is available now, in particular
the long-wavelength L-shell ions of Mg, Si, S and Ar as well as the inner-shell
UTA transitions of M-shell iron ions.

We produced a grid of models using the following ionizing spectrum.  For
$\lambda>80$~\AA\ we used the parameterization of Mathur et al.
(\cite{mathur95}) and below $80$~\AA\ we used the continuum as derived from the
present fit to the Chandra data, with an exponential high-energy cut-off at
$115$~keV as found before using BeppoSAX observations of this source (Nicastro
et al.  \cite{nicastro}).  The $1$--$1000$~Ryd flux of the model is $3.10\times
10^{-13}$~W\,m$^{-2}$, corresponding to an intrinsic source luminosity of
$3.78\times 10^{37}$~W in the same energy band.  We fixed the gas density to
$10^{14}$~m$^{-3}$, but the model is not very sensitive to this.  Abundances
were fixed to solar values (Anders \& Grevesse \cite{anders}), and we used small
column densities in order to prevent strong variations of the ion concentrations
within the slab.  The ionization parameter $\xi\equiv L/nr^2$ is expressed in
the usual units of $10^{-9}$~W\,m (1~erg\,s$^{-1}$\,cm) throughout this paper.
Here $L$ is the ionizing luminosity in the 1--1000~Ryd band, $n$ the gas density
and $r$ the distance of the slab from the ionizing source.

We calculated models for $\log \xi$ ranging from 0.4 to 4.0 in steps of 0.1.
The output of the model consists of a set of ionic column densities.  With these
we tried to explain the observed column densities as a linear combination of a
minimal number of ionization components.  A single component is not sufficient
to explain the broad range of ions. We need at least two additional ionization
components.  We label these components A, B and C with decreasing ionization
parameter $\xi$.

The dominant ionization component (B) produces amongst others the strong
\ion{O}{vii} absorption lines, and was first discovered by Kaastra et al.
(\cite{kaastra2000}).  In addition, we find evidence for a second, less ionized
component (C).  This mainly produces the UV lines, as well as \ion{C}{v},
\ion{Si}{viii} lines and the UTA of \ion{Fe}{vii}--\ion{Fe}{xii}.  Finally, a
very high ionization component (A) producing most of the hydrogenic ions as well
as the L-shell transitions of \ion{Fe}{xix}--\ion{Fe}{xxiv} is also found.

\begin{table}[!h]
\caption{Total column densities, ionization parameters and abundances
for the three components A, B and C.
The listed abundances are the logarithms
of the abundances in solar units.}
\label{tab:xifit}
\centerline{
\begin{tabular}{|l|rrr|}
\hline
Parameter & A  & B  & C \\
\hline
$\log$ $N_{\mathrm H}$ (m$^{-2}$) & 26.93$\pm$0.20 & 25.67$\pm$0.17 & 23.88$\pm$0.13 \\
$\log \xi$                  & 2.87$\pm$0.15 & 1.87$\pm$0.17 & 0.50$\pm$0.20 \\
$T$ (eV) & 43$\pm$8 & 8.0$\pm$2.4 & 2.0$\pm$0.2 \\
\hline
Abundances: &&& \\
C  && -0.08$\pm$0.12 &\\
N  && -0.32$\pm$0.15 &\\
O  && -0.56$\pm$0.07 &\\
Ne && -0.19$\pm$0.16 &\\
Na &&  0.33$\pm$0.38 &\\
Mg && -0.17$\pm$0.18 &\\
Si && -0.32$\pm$0.14 &\\
S  &&  0.03$\pm$0.16 &\\
Ar && -0.21$\pm$0.53 &\\
Fe & -1.18$\pm$0.43 &  0.02$\pm$0.07 & 1.45$\pm$0.56 \\
\hline\noalign{\smallskip}
\end{tabular}
}
\end{table}

In Table~\ref{tab:ioncol} we list in addition to the observed column densities
our best-fit total column densities, as well as the percentage contribution of
each of the three components.  The best fit ionization parameters, column
densities and abundances for each component are listed in Table~\ref{tab:xifit}.
The temperatures listed in this Table are not free fit parameters, but are
derived from the XSTAR models.  In our fit, we have assumed that the abundances
are the same in each component, except for iron, where we allowed the abundance
of each component to vary independently.  In fact, we could not obtain a good
fit if we forced the iron abundance to be the same.  We will discuss these
results further in Sect.~\ref{sect:discussion_wa}.

\subsection{Narrow emission lines}

\begin{figure}
\resizebox{\hsize}{!}{\includegraphics[angle=-90]{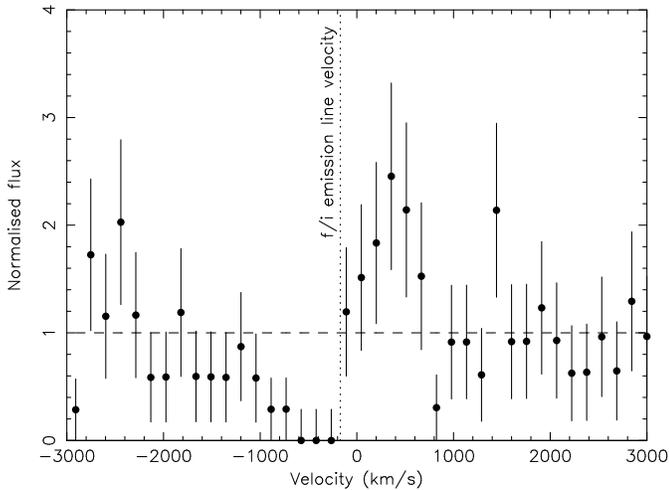}}
\caption{MEG spectrum around the \ion{O}{viii} Ly$\alpha$ line.  Bin size is
$0.01$~\AA, negative velocities indicate blue-shift.  The spectrum has been
normalized to 1 for the continuum.  The average velocity of the forbidden and
intercombination lines of \ion{O}{vii} and \ion{Ne}{ix} is also indicated.}
\label{fig:o8meg}
\end{figure}

\begin{table}[!h]
\caption{Narrow emission lines fluxes. Unit: Photons\,m$^{-2}\,$s$^{-1}$,
corrected
for Galactic absorption. Wavelengths are the rest frame wavelengths in \AA.
}
\label{tab:emlines_f}
\centerline{
\begin{tabular}{|lrrr|}
\hline
line & $\lambda$ & LETGS & MEG  \\
\hline
\ion{O}{viii} Ly$\alpha$ & 18.969 &  0.16$\pm$0.12 &  0.19$\pm$0.10  \\
\ion{O}{vii} r           & 21.602 &  0.30$\pm$0.16 &  0.15$\pm$0.15  \\
\ion{O}{vii} i           & 21.804 &  0.27$\pm$0.17 &  0.20$\pm$0.15  \\
\ion{O}{vii} f           & 22.101 &  0.81$\pm$0.19 &  0.82$\pm$0.26  \\
\ion{Ne}{ix} f           & 13.700 &  0.25$\pm$0.10 &  0.09$\pm$0.04  \\
\hline\noalign{\smallskip}
\end{tabular}
}
\end{table}

\begin{table}[!h]
\caption{Velocities of the narrow emission lines, 
rounded to multiples of $10$~km/s.
The last column gives the MEG values corrected for a velocity of $-172$~km/s
as described in the text.}
\label{tab:emlines_v}
\centerline{
\begin{tabular}{|lrrr|}
\hline
line & LETGS & MEG & MEG (corr.) \\
\hline
\ion{O}{viii} Ly$\alpha$ & +650$\pm$550 & +240$\pm$110 & +420 \\
\ion{O}{vii} r           & +680$\pm$240 & +170$\pm$180 & +340 \\
\ion{O}{vii} i           & +10$\pm$330 & -120$\pm$140 & +50 \\
\ion{O}{vii} f           & -70$\pm$100 & -180$\pm$50 & -0 \\
\ion{Ne}{ix} f           & -330$\pm$260 & -200$\pm$130 & -30 \\
\hline\noalign{\smallskip}
\end{tabular}
}
\end{table}

Only a limited number of narrow emission lines is detected significantly in the
X-ray spectrum of NGC~5548.  These lines are listed in Table~\ref{tab:emlines_f}
and Table~\ref{tab:emlines_v}.  The intercombination ($i$) and forbidden ($f$)
lines of the He-like \ion{O}{vii} and \ion{Ne}{ix} have a very small oscillator
strength, and therefore no significant absorption component.  This makes it
easier to measure the intensity and average velocity than for the
resonance lines.  The \ion{O}{vii} resonance line ($r$) as well as the
\ion{O}{viii} Ly$\alpha$ line have a strong absorption component.  For the LETGS
data, the uncertainties on the average velocities include the systematic
uncertainties of Fig.~\ref{fig:wavacc}, which are much smaller than the
statistical errors.  The Chandra proposers Guide lists a relative wavelength
accuracy for the MEG data of $0.0055$~\AA, and $0.011$~\AA\ in absolute value.
At $22$~\AA, this corresponds to an uncertainty of $75$ and $150$~km/s,
respectively.  From Table~\ref{tab:emlines_v} we see that the intercombination
and forbidden lines as measured with the MEG are all blue-shifted, with an
average value of $-172\pm 47$~km/s.  This is close to the systematic
uncertainty, and significantly different from the centroids as measured with the
LETGS.  In the last column of Table~\ref{tab:emlines_v} we have given the line
centroids with this constant offset of $-172$~km/s subtracted.  With this
correction, both wavelength scales agree within $120$~km/s.

\subsection{Radiative Recombination Continua\label{sect:rrc}}

NGC~5548 produces mostly strong X-ray absorption lines.  The strongest emission
lines visible in the spectrum are the forbidden and intercombination lines of
He-like oxygen and neon.  These lines are mostly formed through recombination
under the conditions that prevail in NGC~5548.  The upper levels of these lines
are populated by radiative cascades following radiative recombination from
shells above $n$=2, as well as by direct recombination into the $n$=2 shell.  A
recombining plasma in photoionization equilibrium should also yield a
significant number of recombinations directly to the ground state.  We thus
expect significant Radiative Recombination Continua (RRC), which due to the
prevailing low temperatures should have a narrow width.

The most promising candidate appears to be the \ion{O}{vii} RRC, given the
strong forbidden line, as well as the large column density of its parent ion
\ion{O}{viii}.  We could not detect a significant {\it narrow} RRC at the
expected wavelength of $16.77$~\AA, and its best-fit absorption-corrected flux
is $0.06\pm 0.10$~photons\,m$^{-2}$\,s$^{-1}$.  Comparing this with the strength
of the forbidden line from Table~\ref{tab:emlines_f} we find a ratio RRC/$f$ of
$0.07\pm 0.12$.  This is much smaller than predicted by recombination models.
For example, Behar et al.  (\cite{behar2001}) in their modeling for the Seyfert
2 galaxy NGC~1068 find a ratio RRC/$f$ of 0.57 for an adopted temperature of
5~eV.  The temperature dependence on this ratio is weak.  Also in the MEG data
this ratio is small:  0.22$\pm$0.25.  Thus, there appears to be a problem with
the predicted strength of the \ion{O}{vii} RRC.

\begin{figure}
\resizebox{\hsize}{!}{\includegraphics[angle=-90]{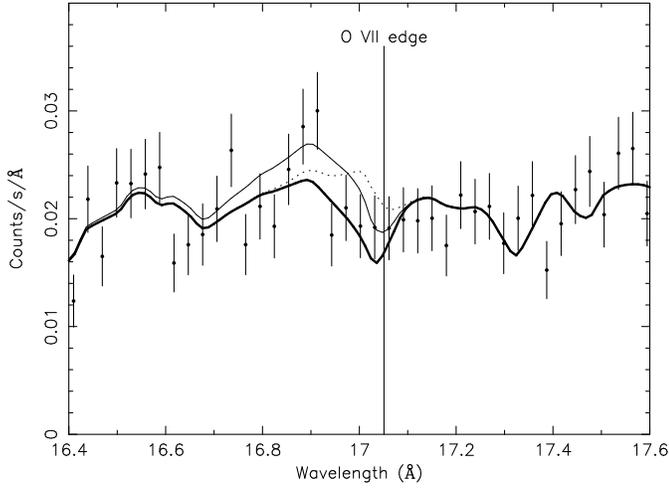}}
\caption{The spectrum near $17$~\AA\ with the best fit model of 
Fig.~\ref{fig:sfit03} only, (thick solid line) or with the \ion{O}{vii}
RRC included for a temperature of 2~eV (dotted line) and 8~eV (thin solid line).
The flux of the RRC was derived from the forbidden line of 
\ion{O}{vii}.}
\label{fig:o7rrc}
\end{figure}

But are we really missing the RRC?  We calculated the flux and spectrum of the
RRC based upon the \ion{O}{vii} forbidden line as a function of temperature
(Fig.~\ref{fig:o7rrc}).  We see from this figure why a low temperature RRC does
not fit:  it over-predicts the flux around $17.0$~\AA.  However for a higher
temperature of 8~eV a better match is obtained, in particular near $16.9$~\AA.
On the other hand, there is now a $\sim 2\sigma$ flux deficit near $16.95$~\AA.
This corresponds to a rest frame wavelength of $16.67$~\AA\ and coincides with
the strongest inner shell transition of \ion{Fe}{viii} (see also
Table~\ref{tab:ioncol}).  Thus the dip can be compensated for in principle by
slightly increasing the column density of \ion{Fe}{viii}.  It is evident that
the situation near this wavelength is rather complex.  This is even more the
case given the fact that one of the strong \ion{Fe}{xvii} lines at $16.78$~\AA\
is also within $0.01$~\AA\ or $200$~km/s from the \ion{O}{vii} edge.  This line
has an estimated optical depth at line center of 1.0 for each velocity
component.  Therefore it will interfere with the RRC (in fact, the dip near
$17.05$~\AA\ in the original model of Fig.~\ref{fig:o7rrc} is due to this
\ion{Fe}{xvii} line).

Given this and the limited statistics of our data further hard conclusions are
difficult to obtain.  We conclude that the data are not inconsistent with RRC
emission of a temperature of 8~eV, the predicted temperature of the \ion{O}{vii}
ions based upon our XSTAR modeling.

\begin{figure}
\resizebox{\hsize}{!}{\includegraphics[angle=-90]{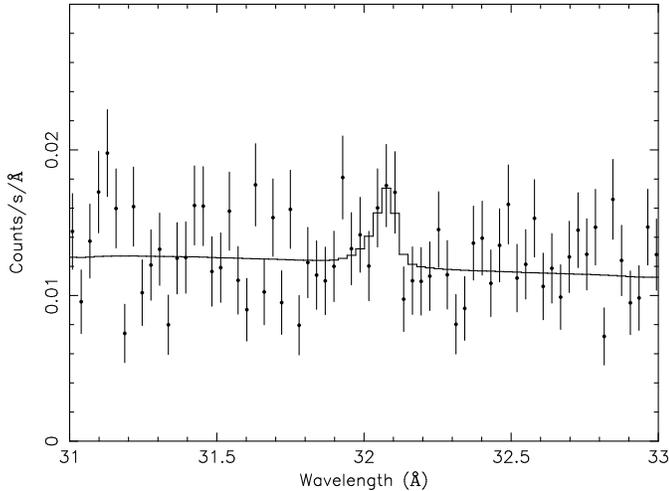}}
\caption{The best-fit \ion{C}{v} RRC superimposed upon the local continuum.}
\label{fig:c5}
\end{figure}

Another important RRC is that of \ion{C}{v}.  There is a clear indication for
the presence of this RRC in the spectrum (see Fig.~\ref{fig:c5}).  We find an
absorption-corrected flux of $0.47_{-0.24}^{+0.45}$~photons\,m$^{-2}$\,s$^{-1}$,
corresponding to an emission measure $n_{\rm e}n_{\rm C\,VI}V$ of $9.5\times
10^{66}$~m$^{-3}$, where $n_{\rm e}$ is the electron density, $n_{\rm C\,VI}$
the density of the parent ion \ion{C}{vi} and $V$ the emitting volume.  For the
temperature we obtain only an upper limit:  the best-fit value is $0.5\pm
0.5$~eV, which is smaller than the modeled temperature of ionization component C
($2.0$~eV).  The corresponding forbidden line of \ion{C}{v} at $41.47$~\AA\ is
difficult to measure, because it is redshifted into the deep neutral carbon edge
region of the LETGS instrument.  Formally, we find a best-fit absorption
corrected flux of $0.26\pm 0.34$~photons\,m$^{-2}$\,s$^{-1}$, but the systematic
uncertainty is likely to be as large as the statistical error.  Hence, the
RRC/$f$ ratio for \ion{C}{v} remains inconclusive.

The measured blue-shift of the \ion{C}{v} RRC corresponds to an outflow velocity
of $-470\pm 240$~km/s for the recombining \ion{C}{vi} ions.  This is in
agreement with the average blue-shift of the \ion{C}{vi} Ly$\alpha$ line
(Fig.~\ref{fig:c6line}).  However for a complete and spherically symmetric,
outflowing shell around the nucleus containing \ion{C}{vi} we would expect to
see no net velocity of the RRC.  The absence of a red (and zero velocity)
component of the RRC can be understood if only a partial shell (opening angle
typically 30--60 degrees) is visible.  This could be due to some degree of
collimation and/or obscuration of the red part by the accretion disk or torus.

\subsection{Low energy emission lines}

In the past there has been a debate whether the EUVE spectrum of NGC~5548
contains emission lines (in particular \ion{Ne}{vii}/\ion{Ne}{viii} at
$88.1$~\AA\ and \ion{Si}{vii} at $70.0$~\AA, Kaastra et al.  \cite{kaastra95})
or that it is a featureless power law (Marshall et al.  \cite{marshall}).  Our
spectral modeling (see Fig.~\ref{fig:sfit09}) clearly shows that at long
wavelengths the spectrum is not a simple power law or other smooth continuum,
but that there are many absorption lines.  In fact, our knowledge of the long
wavelength lines is currently limited.  We expect that many more lines should be
present in this energy band than taken into account here, due to other L-shell
transitions of ions from Ne, Mg, Si, S, Ar and M-shell transitions of iron ions.
Our model only contains the strongest of those lines.

We also tested the alternative hypothesis of (thermal) line emission in addition
to the modified blackbody and power law continuum.  Taking a temperature of
0.06~keV, derived from the EUVE data by Kaastra et al.  (\cite{kaastra95}), we
find a best-fit emission measure of $(4\pm 5)\times 10^{70}$~m$^{-3}$.  This is
only 2~\% of the emission measure found by Kaastra et al.  However for a lower
temperature of 0.025~keV we get an emission measure that is consistent with this
EUVE value.

We conclude that if the lines seen by EUVE are real, then the emitting component
must be time-variable.  Our LETGS spectrum has much better spectral resolution
then EUVE, but due to the low count rates the signal to noise ratio per
resolution element is rather limited.  Therefore it is hard to identify
individual spectral features beyond $50$~\AA.  There are some suggestions for
emission features/complexes, e.g.  near $79$, $82$ and $87$~\AA, but we find no
obvious identifications.  Higher signal to noise data are clearly needed in the
long wavelength band in order to resolve this problem.

\section{Discussion\label{sect:discussion}}

\subsection{Continuum}

The global shape of the continuum (power law plus soft excess, modeled here by a
modified blackbody), is not too different from what has been found in the past,
for example using EXOSAT data (Kaastra \& Barr \cite{kaastra89}).  We will not
discuss this here in detail, but only note that the modified blackbody component
can be attributed to continuum radiation from the accretion disk.  The reason
that we expect a modified blackbody instead of a simple blackbody is that
Compton scattering in the disk cannot be neglected as compared to free-free
absorption.  However, at the relatively low temperatures that we find, the
absorption opacity is not only caused by free-free absorption, but also
free-bound and line opacity are important in the soft X-ray band.  This last
issue points us to the study of relativistic emission lines from the disk.

\subsection{Relativistic oxygen and nitrogen lines}

We have investigated the possible existence of relativistically broadened
emission lines of C, N and O.  Similar very strong lines were discovered by
Branduardi-Raymont et al.  (\cite{branduardi}) in MCG --6-30-15 and Mrk~766.  We
find evidence for such lines (at the 3$\sigma$ confidence level) for
\ion{O}{viii} and \ion{N}{vii}.  However the line intensity in our case is much
weaker, about 20~\% of the underlying continuum.  In particular, we find
equivalent widths of $0.6$ and $1.1$~\AA\ for both lines, an order of magnitude
smaller than the results from the narrow line Seyferts studied by
Branduardi-Raymont et al.  These authors argued that the lines are produced in
the inner, irradiated parts of the accretion disk.  We follow this
interpretation but do not discuss here in detail the formation process of these
lines.  We conclude that apparently the emerging line power in NGC~5548 is much
smaller than in these other sources.  In addition, in NGC~5548 the \ion{C}{vi}
line is not detected.

The inclination angle derived from our fit to the \ion{O}{viii} and \ion{N}{vii}
Ly$\alpha$ profiles is 46 degrees (Table~\ref{tab:contfit}).  This is consistent
with the results of Yaqoob et al.  (\cite{yaqoob}) for the Fe-K line in their
re-analysis of ASCA data including the presence of a narrow emission line as
observed with the HETGS of Chandra.  In the case of the \ion{O}{viii} and
\ion{N}{vii} lines we find a rather steep emissivity law ($q$=4).  The inner
radius of the disk is not very well constrained, but its upper limit of
2.6$GM/c^2$ indicates that the central black hole is more likely of the Kerr
than the Schwarzschild type.

Interestingly, Nicastro et al.  (\cite{nicastro}) in their analysis of BeppoSAX
data of NGC~5548 found evidence for a hard to explain, weak excess emission
feature with an equivalent width of $2.3\pm 1.7$~\AA\ centered around an
observed wavelength of $23\pm 3$~\AA.  Nicastro et a.  tried to explain this by
an unresolved blend of \ion{O}{vii} and \ion{O}{viii} emission lines.  To
explain the high EW measured, these lines were supposed to be produced in a
truly warm gas with a temperature higher than that of photoionization
equilibrium.  Our high-resolution data are incompatible with this explanation.
Instead, the centroid and equivalent width of the feature are consistent with
our current model for the relativistic nitrogen line.

\subsection{The \ion{C}{vi} Ly$\alpha$ line}

We have found no evidence for a relativistic \ion{C}{vi} Ly$\alpha$ line.
However we found a narrower emission component centered around the rest frame
wavelength of this line.  There is not much asymmetry in the line profile.  Its
fitted width $\sigma_{\mathrm v}=4500\pm 1400$~km/s which translates into a FWHM
of $10\,600$~km/s.  This is consistent with the measured width of the optical
broad lines of highly ionized species, such as \ion{C}{iv} and \ion{He}{ii},
which also have widths up to $10\,000$~km/s.  Therefore we propose that the
broad emission component of \ion{C}{vi} Ly$\alpha$ originates from the same
material that also emits the UV/optical broad lines.  We note that the average
photon flux in the broad \ion{C}{iv} line is three orders of magnitude larger
than the photon flux we find for the \ion{C}{vi} line.  From our XSTAR runs we
find for a model with thin irradiated clouds the column density of \ion{C}{vi}
to be 1000 times smaller than the \ion{C}{iv} column density, if the ionization
parameter $\log\xi$ is about -1.  Such an ionization parameter is not
unrealistic for the broad line region.  Interestingly, for this ionization
parameter the \ion{C}{v} column density should be larger than the \ion{C}{vi}
column density.  In principle we might thus also expect a similar emission
component for the strongest \ion{C}{v} lines.  Inspecting the fit residuals of
Fig.~\ref{fig:sfit06} and Fig.~\ref{fig:sfit07} around the \ion{C}{v} 1s$^2$ -
1s3p$^1$P$_1$ and 1s$^2$ - 1s2 $^1$P$_1$ lines, at $34.973$ and $40.268$~\AA,
this is indeed suggested by the data, although for the latter case the
significance is restricted by the poor effective area around $40$~\AA.

\subsection{Warm absorber\label{sect:discussion_wa}}

Our analysis confirms the dominance of a medium ($\xi\sim100$) ionization
component B (Kaastra et al.  \cite{kaastra2000}).  Component B produces most of
the strong absorption lines in the $10$--$25$~\AA\ band.  In addition, we have
found evidence for at least two other components at low and high ionization
parameter, discussed in more detail below.

\subsubsection{Low ionization component C}

Evidence for the low ionization component C arises mainly from the blend of weak
iron inner shell transitions (unresolved transition array or UTA) in the
$16$--$17$~\AA\ band, as well as from the UV lines in NGC~5548.  It is
characterized by an ionization parameter 20 times smaller than for component B,
and has a significantly smaller column density.  How much smaller is hard to
say.  Table~\ref{tab:xifit} shows a problem with the apparent abundance ratio of
iron to the other elements (mainly C, N and O).  Our formal fit in
Table~\ref{tab:xifit} needs an overabundance of iron in the low ionization
component of 1--2 orders of magnitude as compared to the medium ionization
component B.  Component B in general does not show too strong deviations from
solar abundances.

There are reasons to mistrust this apparent abundance anomaly.  In the first
place, it is evident from Table~\ref{tab:ioncol} that the column densities of H,
C, N and O in the low ionization component are all determined from UV
measurements (either FUSE or HST).  These observations were not taken
simultaneously with the current X-ray data.  But even more important than the
non-simultaneity is that these abundances are all derived from strongly
saturated lines (see the last column of Table~\ref{tab:ioncol}).  In this case
the equivalent width of the lines is mainly proportional to the covering factor
and the velocity width $\sigma_{\mathrm v}$; the dependence upon column density
is only logarithmic.  This may easily cause strong bias.  This was shown by Arav
et al.  (\cite{arav}), who demonstrated that the column density of C~IV
increases by a factor of 4 if covering factors are properly taken into account.
However, in our analysis we have accounted for those effects.

On the other hand, the iron abundance of component C is determined completely by
the many weak, unresolved lines of the UTA, all of which have small optical
depth.  Although the UTA is definitely present, the column densities of the
individual ions are sensitive to systematic effects.  Small calibration
uncertainties may still be present at the few percent level.The relevant
wavelength region contains some CsI edges that were taken into account as well
as possible.  Also, our analysis of the \ion{O}{vii} RRC (Sect.~\ref{sect:rrc})
shows that this feature influences the results for the UTA region.  Overall we
estimate that these systematic effects will not affect the strength of the UTA
by more than a factor of two.  They may even increase the UTA's strength.

The weakness of the UTA in the current spectrum makes it hard to determine if
component C is dominated by a single ionization parameter $\xi$, or that it has
multiple contributions.  In the last case, the slightly different sensitivity to
$\xi$ of the UV lines as compared to the iron lines may also cause some
differences.

Finally, the modeling of the UV lines only takes into account the resonance
absorption of UV photons by the relevant ions.  Since the optical depth of the
lines is so large, there should also be a significant emission component.  Given
the complex dynamical structure of the absorber, it is not easy to predict
precisely how this will affect the total equivalent width of the absorption
line.  However if there are significant emission components intermixed with the
absorption, the column density of the UV lines may be under-predicted.

Of course component C could really have a higher iron abundance.  Indications
for a high iron abundance in another AGN have recently been presented by Boller
et al.  (\cite{boller}).  They found a sharp drop in the spectrum above 7.1~keV
in an XMM-Newton observation of the Narrow-Line Seyfert 1 galaxy 1H~0707-495.
In their best model this is interpreted as \ion{Fe}{i} K-shell absorption caused
by partially covering material with a hydrogen column density of a few times
$10^{26}$~m$^{-2}$ and a strong overabundance of iron by a factor of 35.  This
overabundance is similar to what we deduce for component C, however we find a
much higher degree of ionization (around \ion{Fe}{xi}) and two orders of
magnitude smaller column density.

It is evident that spectra with better signal to noise ratio, preferably
simultaneously with UV observations are needed to resolve this problem.

\subsubsection{High ionization component A}

Evidence for the high ionization component A mainly arises from the strong
K-shell resonance absorption lines of S, Si and Mg.  Also a major part of the
hydrogenic \ion{Ne}{x} and \ion{O}{viii} column densities is generated by this
component (see Table~\ref{tab:ioncol}), as well as most of the iron absorption
lines from \ion{Fe}{xix}--\ion{Fe}{xxiii}.  It is puzzling why the apparent iron
abundance of this component is so small:  iron appears in this component to be
under-abundant by a factor of 10 with respect to the other metals.

\begin{table}[!h]
\caption{$\sigma_{\mathrm{th}}$ in km/s for the three components
for hydrogen, oxygen and iron ions.
}
\label{tab:velo}
\centerline{
\begin{tabular}{|l|rrr|}
\hline
Comp.     & A  & B  & C \\
   T (eV) & 43   &  8   &  2 \\
\hline
H         & 64  & 27  & 14 \\
O         & 16  & 6.9 & 3.4 \\
Fe        & 8.6 & 3.7 & 1.8 \\
\hline\noalign{\smallskip}
\end{tabular}
}
\end{table}

A possible explanation is the following.  The strong lines of O, Ne, Mg and Si
(dominating the determination of the column density $N$) have large optical
depths (Table~\ref{tab:ioncol}).  Therefore their equivalent width $EW$ scales
like $\sigma_{\mathrm v} \ln N$, and thus depends only weakly upon $N$.  Most
high ionization iron lines have small optical depths, in which case the opposite
is true:  $EW \sim N$.  Therefore the equivalent width does not depend upon
$\sigma_{\mathrm v}$ but is only proportional to the column density (and thereby
the abundance).  In our analysis we assumed that all lines have the same value
for $\sigma_{\mathrm v}$ as the strong UV components, namely $32$~km/s.
Table~\ref{tab:velo} shows the thermal contribution $\sigma_{\mathrm{th}}$ to
the width for each of the three ionization components.  The total width
$\sigma_{\mathrm v}$ is obtained by combining the thermal width with any
turbulent broadening $\sigma_{\mathrm{turb}}$ that is present:  \begin{equation}
\sigma_{\mathrm v}^2 = \sigma_{\mathrm{th}}^2 + \sigma_{\mathrm{turb}}^2.
\end{equation} If, for instance, component A has a small turbulent velocity
$\sigma_{\mathrm{turb}}$, then $\sigma_{\mathrm v}$ for oxygen can be as small
as $16$~km/s.  If that is the case, we need a column density that is two orders
of magnitude larger for the strongly saturated ions of O, Ne, Mg and Si (since
$\log N\sim EW / \sigma_{\mathrm v}$)).  However for iron we would still obtain
the same column density and thus this would worsen the abundance difference.
Clearly, the opposite must be true and thus component A must have
$\sigma_{\mathrm v}$ in the range of 1--2 times $32$~km/s.  Kaastra et al.
(\cite{kaastra2000}) derived a typical value for $\sigma_{\mathrm v}$ of $140\pm
30$~km/s for \ion{O}{vii}, \ion{O}{viii} and \ion{C}{vi} based upon line ratios
using lines from the same ion.  This was derived under the assumption of a
single velocity component.  Comparing this width to our current estimate, it
follows that we need between 2--5 velocity components each with $\sigma_{\mathrm
v}=$1--2 times $32$~km/s, to reproduce the total effective width of $140$~km/s.

A part of this may be due to the high velocity component 1 ($-1056$~km/s).  In
our analysis we assumed that the ionization component A is dominated by the
velocity components 2--5 ($-669$ to $-163$~km/s).  For the low wavelength lines
we have insufficient spectral resolution to disentangle the velocity components.
For \ion{C}{vi} we clearly separate component 1 from 2--5
(Fig.~\ref{fig:c6line}).  Our model predicts that \ion{C}{vi} contains
contributions from both components A and B.

The total column density of component A, derived from the iron lines is
approximately $5\times 10^{25}$~m$^{-2}$.  This assumes that iron has the same
abundance in component A and B.

\subsubsection{The high velocity component}

The situation around the high velocity component 1 ($v=-1056$~km/s) is rather
complex and confusing.  It was first discovered for \ion{N}{v} and \ion{C}{iv}
lines in Goddard High-Resolution Spectrograph (GHRS) observations taken in
August 1996 (Crenshaw et al.  \cite{crenshaw}), and for \ion{H}{i} Ly$\alpha$
and \ion{N}{v} in March 1998 in STIS spectra (Crenshaw \& Kraemer
\cite{crenshawk}).  The dynamical structure (velocities, widths) did not change
over these two years, although the column density decreased by a factor of 4.
Most other velocity components (2--5) did not change that much (except component
3 at $-540$~km/s which decreased by $\sim$40~\%; this component is the most
highly ionized of components 2--5).

FUSE observed NGC~5548 in June 2000 during a low state (Brotherton et al.
\cite{brotherton}).  It reconfirmed the presence of component 1 in the
\ion{O}{vi} and \ion{H}{i} Ly$\beta$ lines.  The \ion{H}{i} column reported by
Brotherton et al.  however is 6 times larger than the column reported by
Crenshaw \& Kraemer which was based upon the Ly$\alpha$ observation two years
earlier.  In Sect.~\ref{sect:fitslab} we have seen that a reanalysis of Arav et
al.  (\cite{arav}) for the \ion{C}{iv} line led to a four times larger column
density for component 4.  Since the optical depth of the hydrogen lines is even
larger (Table~\ref{tab:ioncol}), we suppose that the Crenshaw \& Kraemer
\ion{H}{i} column density was underestimated.

The column densities of \ion{O}{vi} and \ion{H}{i} Ly$\beta$ for the dynamical
component 1 as derived from the FUSE data, are consistent with $\log \xi =
2.35\pm 0.06$ and a hydrogen column density of $1.0\times 10^{26}$~m$^{-2}$, if
modeled by XSTAR.  This column density is ten times higher than the column
derived by Brotherton et al., on the other hand we also have a slightly larger
ionization parameter than these authors.  For large $\xi$ the \ion{O}{vi} and
\ion{H}{i} column densities are very sensitive to the details of the
photoionization model used.  The same model predicts a \ion{Si}{xiv} column of
$\log N$ = 20.83, smaller than the total observed column of
$22.19_{-0.85}^{+0.36}$.  But the observed value has (1) a large statistical
error due to strong line saturation and (2) it contains contributions from all
velocity components.  Similar results are obtained for other highly ionized ions
such as \ion{Mg}{xii}.  For \ion{C}{vi} the model predicts $\log N = 21.30$
(compared to the measured value of $21.32\pm 0.31$ for the components 2--5, cf.
Table~\ref{tab:ioncol}).  Given that component 1 of \ion{C}{vi} contains about a
third of the total equivalent line width of the line (Table~\ref{tab:c6line}),
the deviations are not too large.  However, the model under-predicts the
\ion{N}{v} column by an order of magnitude.

Thus, component 1 has a high ionization parameter.  We cannot exclude, however,
that it also has a component with a smaller ionization parameter (needed for
\ion{N}{v} and \ion{C}{iv}), or that it is strongly time variable.  It is
evident that simultaneous X-ray/UV measurements with high S/N are needed to
resolve this problem.

\subsubsection{Comparison with previous X-ray observations}

It is not easy to compare the properties of the warm absorber as we derived it
here from absorption lines in high resolution LETGS spectra with results
obtained by low resolution instruments.  Fig.~\ref{fig:sfit03} illustrates that
the wavelength-averaged transmission of the warm absorber in the $12$--$17$~\AA\
band is typically between 0.7--0.8.  However the depth of the continuum
absorption edge of \ion{O}{vii} ($16.77$~\AA) is only 9~\% and for \ion{O}{viii}
($14.23$~\AA) 15~\%.  Effectively, these edges are enhanced by the blends of the
UTA from M-shell iron ions and \ion{O}{viii} Ly$\beta$ (for \ion{O}{vii}); by
blends from L-shell iron ions and neon lines (for \ion{O}{viii}).

This may explain why Reynolds (\cite{reynolds}) finds edges of 22~\%
(\ion{O}{vii}) and 15~\% (\ion{O}{viii}) in his ASCA data of NGC~5548, somewhat
higher than our values.  The present results for the dominant ionization
component B are in good agreement with previous studies using ASCA (George et
al.  \cite{george1998}) and BeppoSAX (Nicastro et al.  \cite{nicastro}).  This
is demonstrated in Table~\ref{tab:xicomp}.  Ionization parameters used by those
authors were converted to our $\xi$ convention using the continuum spectral
shape of Sect.~\ref{sect:modcol}.  Both the column densities and the ionization
parameters are consistent with the values presented here.  It should be noted
that the 2--10~keV fluxes during all these observations were rather similar.
The HETGS data (taken when the source was two times fainter) also show no
evidence for large differences in the warm absorber properties, indicating that
component B is rather stable.

\begin{table}[!h]
\caption{Comparison of the dominant ionization component B}
\label{tab:xicomp}
\centerline{
\begin{tabular}{|lrrl|}
\hline
Instr. & $F_{\displaystyle 2-10\,{\mathrm {keV}}}$ &
           $\log$ $N_{\mathrm H}$ & $\log \xi$  \\
   & (W\,m$^{-2}$) & (m$^{-2}$) & ($10^{-9}$~W\,m) \\
\hline
ASCA$^{\mathrm{a}}$     & 4.3$\times$10$^{-14}$ & 25.71$\pm$0.05 & 1.54$\pm$0.03  \\
ASCA$^{\mathrm{b}}$     & 5.2$\times$10$^{-14}$ & 25.51$\pm$0.09 & 1.90$\pm$0.09  \\
BeppoSAX$^{\mathrm{c}}$ & 3.5$\times$10$^{-14}$ & 25.44$\pm$0.12 & 2.11$\pm$0.16  \\
LETGS$^{\mathrm{d}}$    & 4.0$\times$10$^{-14}$ & 25.67$\pm$0.17 & 1.87$\pm$0.17  \\
\hline\noalign{\smallskip}
\end{tabular}
}
\begin{list}{}{}
\item[$^{\mathrm{a}}$] Reynolds (\cite{reynolds})
\item[$^{\mathrm{b}}$] George et al. (\cite{george1998})
\item[$^{\mathrm{c}}$] Nicastro et al. (\cite{nicastro})
\item[$^{\mathrm{d}}$] present work
\end{list}
\end{table}

\subsection{Narrow emission lines}

\subsubsection{Intercombination and forbidden lines}

The intercombination and forbidden lines, in the AGN environment formed mainly
by recombination, have almost no net outflow velocity.  This is in contrast to
the strongest $n=2$ to $n=1$ resonance lines from \ion{O}{vii} and
\ion{O}{viii}.  These lines have a strong, blue-shifted absorption component
with velocities of the order of $-500$~km/s, and a redshifted emission component
with a velocity of $+500$~km/s.  The blue emission component of these resonance
lines cannot originate in the same recombining plasma that produces the
non-shifted forbidden and intercombination line.  The discussion given by
Kaastra et al.  (\cite{kaastra2000}) for these last two lines still holds, i.e.
from the low value of $i/f$ an upper limit to the density of $7\times
10^{16}$~m$^{-3}$ can be derived.  Taking the observed flux of the blue
component of the \ion{O}{vii} resonance line as a typical upper limit to the
non-shifted flux of this line, the ratio $G=(i+f)/r$ has a lower limit of 3.6.
This value is expected for photo-ionized plasmas and is significantly larger
than the ratio for collisional plasmas.

The error bars on the derived fluxes are quite large (the lines are rather
weak).  For the strongest line (the \ion{O}{vii} forbidden line), the intensity
did not change much between the LETGS and MEG observations, despite the factor
of 2 lower continuum flux in the latter case.  This, combined with the low
velocity of these lines could indicate that they are formed at a rather large
distance from the central ionizing source.  Other evidence for the large
distance of this component is obtained from the width of the forbidden
\ion{O}{vii} line.  The line is narrow and consistent with no Doppler
broadening.  The upper limit to the intrinsic Gaussian line width
$\sigma_{\mathrm v}$ is $310$~km/s for the LETGS data and 330~km/s for the MEG
data.  This would place this component at the same distance as the optical
narrow line region.

\subsection{Resonance lines}

The \ion{O}{vii} resonance and \ion{O}{viii} Ly$\alpha$ line profiles may be
considered as a P Cygni type profile of an absorbing wind.  The optical depth in
the line center of the absorption components of these lines is very large (41
and 85 for \ion{O}{viii} and \ion{O}{vii}, respectively, cf.
Table~\ref{tab:ioncol}).  This causes the resonantly scattered photons to
produce a strong emission component.  The fact that we see a red component means
that we can look behind the position of the X-ray continuum source in the plane
of the sky (we see the receding component of the wind).  This implies that the
scattering region must be located at a radial distance larger than the outer
edge of the (presumably optically thick) accretion disk.

Where is the X-ray absorber located?  We first note that from a comparison of
the X-ray spectra obtained by LETGS and HETGS it is difficult to establish
whether the X-ray absorber varies in time, the limitation being mainly
statistics.  However from the dynamical association with the UV absorbers (in
X-rays we see both the high velocity component 1 as well as a blend of absorbers
at the same velocities as UV components 2--5) we assume that the X-ray absorber
has a similar distance to the central source as the UV absorber.  Arav et al.
(\cite{arav}) have argued that the dominant component of the \ion{C}{iv} line
must be produced somewhere in between the broad and narrow optical line region.

The narrow optical line region in NGC~5548 consists of a low ionization
component at about 70~pc distance plus a higher ionization component at a
distance of 1~pc ($\log U =-1.5$, corresponding to $\log\xi=-0.1$ for our
assumed spectrum).  This last component is still less ionized than our component
C but produces most of the narrow UV emission lines (Kraemer et al.
\cite{kraemer}).  Thus, the warm absorber must be within a distance of about
1~pc.

The high velocity component 1 is strongly variable on the timescale of 1.5 years
(Crenshaw \& Kraemer \cite{crenshawk}).  This puts an upper limit of about 1
light year ($10^{16}$~m) to the size of this component.

Koratkar \& Gaskell (\cite{koratkar}) estimated the luminosity-weighted radius
of the broad \ion{C}{iv} emission line to be 31 light days.  Similar size were
obtained by Dietrich, \& Kollatschny (\cite{dietrich}).  Using a refined
reverberation technique, Wanders et al.  (\cite{wanders}) found that \ion{C}{iv}
should be distributed between 5 and 12 light days.

Thus we argue that the absorber is located between 12 light days and 1 light
year, i.e.  3$\times$10$^{14}$--$10^{16}$~m.  Since we have shown that the outer
radius of the accretion disk must be smaller than the distance to the absorber,
the disk must be truncated at a radius less than a light year.

If the wind is launched from the accretion disk, we expect that the gas has a
significant rotational velocity component.  We see only a redshifted emission
component of the resonance lines, and no sign for rotational broadening larger
than a few hundred km/s.  This indicates that the gas must somehow have lost
much of its initial rotational velocity components.

\section{Summary and conclusions.}

We have shown here the power of high resolution X-ray spectroscopy for active
galactic nuclei.  The warm absorber in NGC~5548 consists of at least three
ionization components, labeled here A ($\xi=1000$), B ($\xi=100$) and C
($\xi=3$) in decreasing ionization order.  The hydrogen column densities of the
first two components are $5\times 10^{25}$~m$^{-2}$, and a comparison of the
parameters for component B with published low-resolution data obtained by ASCA
and BeppoSAX shows that it is not strongly variable.  Component C has a total
hydrogen column smaller than the other two.

In the UV lines at least six velocity components have been identified in the
past.  The X-ray lines span the full dynamical range of these components, but
due to the limited spectral resolution of the LETGS combined with the present
statistics and the lack of simultaneous UV measurements, the precise connection
between velocity components 1--6 and ionization components A--C is difficult to
quantify.  The highest velocity component 1 ($-1056$~km/s) has at least also a
high ionization contribution.

The formal iron abundance of component A is an order of magnitude too small and
that of component C an order of magnitude to large as compared to the main
component B.  We argue that a part of this discrepancy may be due to the strong
saturation of the absorption lines of the reference ions against which we
measure the iron abundance of components A and C.  The relative abundances of
the elements as measured in particular for the strong component B are close to
solar.  Sodium appears to be over-abundant by a factor of two, while N, O and Si
are slightly under-abundant.

The strongest absorption lines (the 1s--2p lines of hydrogenic and helium-like
oxygen) have significant red emission components, with velocities in the range
of $+300$--$+600$~km/s.  Since these velocities are similar to the average
velocity of the blue-ward absorption component, it is evident that we see a
major part of the outflowing wind.  This means that the outer radius of the
accretion disk must be significantly smaller than the typical size of the
absorber, for otherwise the disk would occult the red-ward wing of the
resonantly scattered oxygen line photons.  Combining this with lower and upper
bounds for the distance of the absorber, this outer edge must then be smaller
than a light year.

The forbidden and intercombination lines of helium-like oxygen and neon are most
likely produces in a region much farther out in the active nucleus, perhaps in
or beyond the optical narrow line region.  Arguments that support this is the
almost zero outflow velocity ($-70\pm 100$~km/s for the \ion{O}{vii} forbidden
line), the narrow line width (smaller than $310$~km/s), as well as the fact that
the line intensity did not vary significantly in the two months between the
LETGS and the HETGS observations, despite the factor of two smaller continuum
flux during the latter observation.  The same component most likely produces a
weak \ion{O}{vii} radiative recombination continuum, and possibly also an RRC
from \ion{C}{v}.

We have found evidence for the presence of weak relativistic Ly$\alpha$ emission
lines from \ion{O}{viii} and \ion{N}{vii}.  These lines are statistically
significant but their equivalent width is an order of magnitude smaller than in
the narrow line Seyfert galaxies studied by Branduardi-Raymont et al.
\cite{branduardi}.  The lines are presumably produced in the inner regions of
the accretion disk, for which we derive a typical inclination angle of 46
degrees.  The inner radius of the disk indicates that the black hole is probably
of the Kerr type.  The nitrogen line was probably also detected (but not
identified) in the BeppoSAX data as published by Nicastro et al.
(\cite{nicastro}).

Further out in the nucleus we find evidence for a broadened (FWHM
$10\,000$~km/s) emission line of \ion{C}{vi}.  We propose that this line is
produced by the broad line region in the same material that also produces the
broad component of the UV \ion{C}{iv} line.  This is suggested by the similar
line width and the apparent \ion{C}{vi} to \ion{C}{iv} line ratio.

Thus, we have encountered in NGC~5548 at least four types of emission lines:
\begin{itemize} \item relativistically broadened emission lines from the
accretion disk (\ion{O}{viii} and \ion{N}{vii} Ly$\alpha$); \item a highly
ionized broad line region line (\ion{C}{vi} Ly$\alpha$); \item P Cygni type
emission components from the warm absorber (\ion{O}{viii} Ly$\alpha$ and
\ion{O}{vii} resonance); \item forbidden and intercombination lines from the
narrow line region or beyond (\ion{O}{vii}, \ion{Ne}{ix}).  \end{itemize}

\begin{acknowledgements}
The Laboratory for Space Research Utrecht is supported
financially by NWO, the Netherlands Organization for Scientific
Research. J.S. Kaastra and D.A. Liedahl acknowledge a NATO Collaborative
Research Grant CRG\,960027.
Work at LLNL was performed under the auspices of the U.S.
Department of Energy by the University of California Lawrence Livermore
National Laboratory under contract No. W-7405-Eng-48.
\end{acknowledgements}

\end{document}